\documentclass[aps,prb,floatfix,amsmath,amssymb,preprint,eqsecnum,nofootinbib]{revtex4}
\usepackage[dvips]{graphicx}% Include figure files
\usepackage{dcolumn}% Align table columns on decimal point
\usepackage{bm}% bold math
\usepackage{setspace}   % controllabel line spacing
\newcommand{\ba}{\begin{array}{ccc}}
\newcommand{\ea}{\end{array}}
\newcommand{\nn}{\nonumber}
 \renewcommand{\d}{\partial}

\def\<{\langle}
\def\>{\rangle}
\usepackage{amsmath}
\usepackage{amssymb}
\usepackage{feynmp}
\begin{document}
\title{Impurity spin textures across conventional and deconfined quantum critical points of
two-dimensional antiferromagnets}

 \author{Max~A.~Metlitski}
 \email{mmetlits@fas.harvard.edu}
 \affiliation{Department of Physics, Harvard University, Cambridge MA 02138, USA}

 \author{Subir Sachdev}
 \email{subir_sachdev@harvard.edu}
\affiliation{Department of Physics, Harvard University, Cambridge MA
02138, USA}

\date{\today\\[24pt]}

\begin{abstract}
We describe the spin distribution in the vicinity of a non-magnetic
impurity in a two-dimensional antiferromagnet undergoing a
transition from a magnetically ordered N\'eel state to a paramagnet
with a spin gap. The quantum critical ground state in a finite
system has total spin $S=1/2$ (if the system without the impurity
had an even number of $S=1/2$ spins), and recent numerical studies
in a double layer antiferromagnet (K.~H.~H\"oglund {\em et al.},
Phys. Rev. Lett. {\bf 98}, 087203 (2007)) have shown that the spin
has a universal spatial form delocalized across the entire sample.
We present the field theory describing the uniform and staggered
magnetizations in this spin texture for two classes of
antiferromagnets: ({\em i\/}) the transition from a N\'eel state to
a paramagnet with local spin singlets, in models with an even number
of $S=1/2$ spins per unit cell, which are described by a O(3)
Landau-Ginzburg-Wilson field theory; and ({\em ii\/}) the transition
from a N\'eel state to a valence bond solid, in antiferromagnets
with a single $S=1/2$ spin per unit cell, which are described by a
``deconfined'' field theory of spinons.
\end{abstract}

\maketitle

\section{Introduction}
\label{sec:intro}

There have been  many experimental studies of non-magnetic Zn
impurities substituting for the spin $S=1/2$ Cu ions in spin-gap and
superconducting compounds
\cite{cuge,bobroff,ouazi,yazdani,pan,vajk}. These have stimulated
many theoretical studies of the spin dynamics in the vicinity of a
vacancy ({\em i.e.\/} a site with no spin) in $S=1/2$ square lattice
antiferromagnets
\cite{nagaosa,sandvik1,science,qimp1,sushkov,neto,troyer,sandvik2,qimp2,luscher,syro,sandvik3,sandvik4}.

An important feature of the impurity-response
escaped\footnote{Section II.B.3 of Ref.~\onlinecite{qimp2} contains
results which can be used to extract the spin textures in zero
field.} theoretical attention until recently \cite{sandvik3}.
Consider the regime where the bulk antiferromagnet preserves global
rotational symmetry and has a $S=0$ ground state. Such states can be
reached by deforming the nearest-neighbor antiferromagnet into a
coupled-ladder or coupled-dimer antiferromagnet
\cite{kotov,matsumoto}, in a double-layer antiferromagnet
\cite{wang}, or by adding additional ring-exchange interactions
while preserving full square lattice symmetry\cite{sandvik5}. Now
remove a single $S=1/2$ spin in a system with an even number of
spins, leaving an antiferromagnet with a vacancy and an odd number
of $S=1/2$ spins. We expect this antiferromagnet to have a
doubly-degenerate ground state with total spin $S=1/2$. Without loss
of generality, we can examine the ground state with spin-projection
$S_z = 1/2$. In such a state, even though there is no broken
symmetry and no applied magnetic field (the Hamiltonian has full
SU(2) spin symmetry), the expectation values of the spin projection
on the site $i$, $\langle S_{zi} \rangle$, is non-zero on all $i$
for any finite system of size $L$. The question of interest in this
paper is the following: What is the spatial form of $\langle S_{zi}
\rangle$ ? It is possible that the $S=1/2$ magnetization is pushed
out to the boundaries of the system, far from the impurity: in this
case, it will not be relevant to the impurity properties in the
limit $L \rightarrow \infty$. However, we will find this is not the
case for the antiferromagnets examined in this paper. For the
spin-gap antiferromagnets we consider, the $S=1/2$ magnetization is
bound to the impurity over a length scale inversely proportional to
the spin gap. At the quantum critical points separating the spin gap
states from the N\'eel state, which define `algebraic spin liquids',
we will find, as in Ref.~\onlinecite{sandvik3}, that the impurity
magnetization is delocalized over the entire system, forming a spin
texture with a universal spatial form determined only by the system
size $L$.

First, let us consider the models which have been numerically
studied in Ref.~\onlinecite{sandvik3}. These are antiferromagnets
which have an even number of $S=1/2$ spins per unit cell (such as
the coupled-dimer\cite{kotov,matsumoto} or double layer\cite{wang}
models), which exhibit a transition between a N\'eel state and a
simple spin gap state; the latter state is adiabatically connected
to a state in which the spins in each unit cell are separately
locked into singlets, with negligible resonance between unit cells.
This is a `conventional' transition, described by a
Landau-Ginzburg-Wilson (LGW) theory. A convenient description of
both phases and the quantum phase transition is provided by the O(3)
non-linear sigma model, expressed in terms of a unit vector field
${\bf n}(\vec{x},\tau)$ representing the local orientation of the
N\'eel order parameter. Here $\vec{x}$ is the two-dimensional
spatial position, $\tau$ is imaginary time, and ${\bf n}^2 = 1$
everywhere in spacetime. The bulk action in the absence of the
impurity is the O(3) non-linear sigma model
\begin{equation}
\mathcal{S}_{\rm b}^{\bf n} = \frac{1}{2 g} \int d\tau \int d^2 x
(\partial_\mu {\bf n} )^2 \,, \label{sbn}
\end{equation}
where $g$ is the coupling constant which tunes the antiferromagnet
from the N\'eel state ($g<g_c$) to the spin gap state ($g>g_c$),
$\mu$ is a 3-dimensional spacetime index and a spin-wave velocity
has been set to unity. In this formulation, the influence of the
impurity is represented universally by the following Berry phase
term alone\cite{qimp2} (provided the antiferromagnet is not too far
from the critical point)
\begin{equation}
\mathcal{S}_{\rm imp}^{\bf n} = i S \int d\tau {\bf A} [ {\bf n}
(0,\tau)] \cdot \frac{d {\bf n} (0, \tau)}{d \tau}, \label{simpn}
\end{equation}
for a spin $S=1/2$ antiferromagnet, where ${\bf A}$ is the Dirac
monopole function in spin space with $\nabla_{{\bf n}} \times {\bf
A} = {\bf n}$. Note that $\mathcal{S}_{\rm imp}^{\bf n}$ does not
include any coupling constants, and it depends upon the value of
${\bf n}$ only at $\vec{x}=0$, which is the position of the
impurity.

Now we need to describe the $S=1/2$ ground state of
$\mathcal{S}_b^{\bf n} + \mathcal{S}_{\rm imp}^{\bf n}$ for $g \geq
g_c$. First, we need a proper discussion of the rotationally
invariant $S=0$ ground state without the impurity. While it may be
possible to do this within the context of a small $g$ expansion of
the O(3) non-linear sigma model, the procedure is quite cumbersome
and delicate, requiring a global average over all possible locally
ordered states. We shall instead follow a simpler procedure which is
described in more detail in Section~\ref{sec:on}: we use an
alternative soft-spin, LGW formulation of $\mathcal{S}_b^{\bf n}$ in
terms of a vector order parameter, $\bm{\phi}$, whose length is
unconstrained. The $\bm{\phi} = 0$ saddle point then is an
appropriate starting point for describing the physics of the $S=0$
ground state of the bulk theory and its excitations. Next, we
include the impurity term described by $\mathcal{S}_{\rm imp}^{\bf
n}$, and also apply an infinitesimal magnetic field in the $z$
direction. As we will show in Section~\ref{sec:on}, the Berry phase
effectively localizes the order parameter at the impurity site,
${\bf n} (\vec{x}=0,\tau)$, to a specific orientation on the unit
sphere; in particular, for the $S_z=1/2$ state chosen by the applied
field, we may perform an expansion about a saddle point with ${\bf
n} (\vec{x}=0,\tau) = (1,0,0)$. This expansion quantizes, at each
order, the total spin at $S_z = 1/2$: this was established in
Section~II.C.2 of Ref.~\onlinecite{qimp1} for $g<g_c$, and the same
result also applies here for $g \geq g_c$. The infinitesimal
magnetic field is set to zero at the end, but the spin density of
the $S_z=1/2$ state remains non-zero in this limit.

The results in Section~\ref{sec:on} provide an explicit analytic
realization for the scaling forms presented in
Ref.~\onlinecite{sandvik3} for the spin texture near the impurity.
For the magnetization density, ${\bf Q}$, which is the conserved
Noether ``charge'' density associated with the O(3) symmetry of the
antiferromagnet, we have at $g=g_c$ and zero temperature ($T$) and
in the $S_z = 1/2$ state:
\begin{equation}
\left\langle Q_z (\vec{x}) \right\rangle = \frac{1}{L^2} \Phi_Q
\left( \frac{\vec{x}}{L} \right) \label{Qscale}
\end{equation}
where $\Phi_Q (\vec{r})$ is a universal function obeying the
quantized total spin condition
\begin{equation}
\int d^2 r \Phi_Q (\vec{r} ) = S \,.
\end{equation}
Similarly, the staggered magnetization associate with the N\'eel
order parameter obeys the scaling form
\begin{equation}
\left\langle n_z (\vec{x}) \right\rangle = \frac{1}{L^{(1+\eta)/2}}
\Phi_n \left( \frac{\vec{x}}{L} \right) \label{nscale}
\end{equation}
at $g=g_c$, where $\Phi_n (\vec{r})$ is another universal function,
but its overall scale is non-universal. The exponent $\eta$ is the
anomalous dimension of ${\bf n}$ at $g=g_c$ in the absence of the
impurity.

Let us now turn to the more interesting and much more subtle case of
a ``deconfined'' critical point \cite{senthil1,senthil2}. Here we
are considering antiferromagnets with a single $S=1/2$ spin per unit
cell, and so there is no simple spin-gap state with local singlets.
For the models studied in
Refs.~\onlinecite{senthil1,senthil2,sandvik5}, the spin gap state
has singlet valence bonds which crystallize into a regular
arrangement, breaking the space group symmetry of the square
lattice, while preserving spin rotation invariance. Such a state is
a valence bond solid (VBS), and we are now interested in the
impurity response across the N\'eel-VBS transition. As argued in
Refs.~\onlinecite{senthil1,senthil2} the vicinity of this quantum
critical point is described by the $\mathbb{CP}^1$ field theory
which is expressed in terms of a bosonic `spinon' represented by a
complex spinor field $z_{\alpha} (\vec{x},\tau)$, where $\alpha =
\uparrow, \downarrow$, and the constraint $\sum_\alpha |z_\alpha |^2
= 1$ is obeyed everywhere in spacetime. The N\'eel order parameter,
${\bf n}$ is related to $z_\alpha$ by
\begin{equation}
{\bf n} = z_\alpha^\dagger \vec{\sigma}_{\alpha\beta} z_\beta \,,
\label{nzz}
\end{equation}
where $\vec{\sigma}$ are the Pauli matrices. Also, in our analysis,
we find it useful to generalize to the $\mathbb{CP}^{N-1}$ model
with SU($N$) symmetry, where $\alpha = 1 \ldots N$, and then the
Pauli matrices are replaced by the generators of SU($N$). The action
of the $\mathbb{CP}^{N-1}$ model also involves a non-compact U(1)
gauge field $A_\mu$, and is given by
\begin{equation}
\mathcal{S}_{b}^z =  \int d\tau \int d^2 x \left[ \frac{1}{g}
|(\partial_\mu -i A_\mu )z_\alpha|^2 + \frac{1}{2 e^2}
(\epsilon_{\mu\nu\lambda} \partial_\nu A_\lambda)^2 \right]\,.
\label{sbz}
\end{equation}
This theory describes a N\'eel-ordered phase for $g<g_c$, and a
spin-gap state with VBS order for $g \geq g_c$ (additional Berry
phase terms are needed to obtain the four-fold square-lattice
symmetry of the VBS order\cite{rs}). It is crucial to note that,
unlike the situation in 1+1 dimensions \cite{dadda,witten}, the
models $\mathcal{S}_b^{\bf n}$ (in Eq.~(\ref{sbn})) and
$\mathcal{S}_b^z$ are {\em not\/} equivalent to each other in 2+1
dimensions. This was established in Ref.~\onlinecite{ashvin}, and is
a consequence of the proliferation of `hedgehog' or `monopole'
defects at the critical point of $\mathcal{S}_b^{\bf n}$; such
defects are absent in the $\mathcal{S}_b^z$ theory.

Now let us add an impurity to the field theory in Eq.~(\ref{sbz}).
It was argued in Ref.~\onlinecite{ImpurityKSBC} that the impurity is
now represented by a source term for a static charge $Q=2S$ at
$\vec{x}=0$. Thus
\begin{equation}
\mathcal{S}_{\rm imp}^z = i Q \int d \tau A_\tau (\vec{x} =0, \tau)
 \label{simpz}
\end{equation}
As before, we are now interested in describing the ground state of
$\mathcal{S}_b^z + \mathcal{S}_{\rm imp}^z$, which we expect carries
total spin $S=1/2$. However, now the projection onto the state with
$S=1/2$ cannot be done by the method used for the LGW theory. For
$g\geq g_c$, we begin with a $S=0$ ground state of
$\mathcal{S}_b^z$, but now don't find that the impurity term in
Eq.~(\ref{simpz}) introduces any net spin: the total spin remains at
$S=0$ to all orders in perturbation theory. Clearly, we need the
impurity charge $Q$ to non-perturbatively bind a $S=1/2$ $z_\alpha$
spinon. For $g>g_c$, such binding can be addressed via a
non-relativistic Schr\"odinger equation \cite{ribhu}, the analysis
does not appear appropriate at the main point of interest, $g=g_c$,
where we have a conformal field theory (CFT) with no sharp
quasiparticle excitations. Here we expect the spinon to be smeared
over the whole system of size of $L$. We shall describe this spinon
state by explicitly beginning with a $S=1/2$ state of
$\mathcal{S}_b^z$ and then perturbatively examining the influence of
$\mathcal{S}_{\rm imp}^z$: this is expected to yield correlations in
the true $S=1/2$ ground state of $\mathcal{S}_{b}^z+\mathcal{S}_{\rm
imp}^z$.

Using the language of general SU($N$), let the ground states of
$\mathcal{S}_{b}^z+\mathcal{S}_{\rm imp}^z$ be $|\alpha \rangle$;
these transform under the fundamental representation of SU($N$). To
find the matrix element of some operator $O(\vec{x})$ between states
$|\alpha\rangle$ and  $|\beta\rangle$ of the SU($N$) multiplet, we
compute, \begin{equation} \label{matel} \langle
\alpha|O(\vec{x})|\beta\rangle = \lim_{\mathcal{\mathcal{T}}\to
\infty}\frac{\left\langle z_\alpha(0,\mathcal{T}/2) \exp \left( -i
\int_{-\mathcal{T}/2}^{\mathcal{T}/2} A_\tau (0,\tau) d\tau \right)
O(\vec{x},0)
z^{\dagger}_{\beta}(0,-\mathcal{T}/2)\right\rangle_{\mathcal{S}_b^z}}{\left\langle
z_\alpha(0,\mathcal{T}/2) \exp \left( -i
\int_{-\mathcal{T}/2}^{\mathcal{T}/2} A_\tau (0,\tau) d\tau \right)
z^{\dagger}_{\alpha}(0,-\mathcal{T}/2)\right\rangle_{\mathcal{S}_b^z}}\,.
\end{equation} Effectively, we start with external charge free vacuum, and
then at time $\tau = -\mathcal{\mathcal{T}}/2$ create a spinon
together with the Wilson line, the latter representing the effect of
the external charge $Q=1$. We wait for a long time $\mathcal{T}/2$
to single out the lowest energy state with the quantum numbers of
the operator $z^{\dagger}_{\alpha}$. We then measure the operator
$O(\vec{x})$, again wait time $\mathcal{T}/2$ and annihilate our
spinon together with the external charge. The denominator in
Eq.~(\ref{matel}) serves to cancel out the matrix element for
creating the spinon\,-\,external charge bound state out of the
vacuum (no sum over $\alpha$ is implied in the denominator).
Expressions of type (\ref{matel}) are common when studying the
properties of heavy-light mesons in quantum chromodynamics.

The time $\mathcal{T}$ must be much larger than the gap between
states with the quantum numbers that we are studying. In the spin
gap phase, $g>g_c$, this gap is finite in the infinite volume limit.
However, at the critical point the gap will be of order $1/L$. So
one has to choose $\mathcal{T} \gg L$. Although unusual, this
condition can always be satisfied as we work at zero temperature.

To discuss higher charge impurity ($Q > 1$) one needs to act on the
vacuum with higher $U(1)$ charge composite operators of the $z$
field. The resulting states can form higher representations of
SU($N$) symmetry. For simplicity, we limit ourselves to $Q = 1$
below.

Details of our evaluation of Eq.~(\ref{matel}) in the $1/N$
expansion appear in Section~\ref{sec:finite}. We will obtain results
for the scaling functions appearing in Eq.~(\ref{Qscale}) and
(\ref{nscale}) describing the spin distribution at the deconfined
quantum critical point.

In addition, in Section~\ref{sec:neel} we compute the uniform and
staggered spin distributions in the N\'eel phase of the
$\mathbb{CP}^{N-1}$ model. We find that the short distance behaviour
of spin distributions both at the critical point and in the Neel
phase is in agreement with the impurity scaling theory postulated in
Ref.~\onlinecite{ImpurityKSBC}. In particular, we obtain substantial
additional evidence that the uniform and staggered spin operators
flow to the same impurity spin operator upon approaching the
impurity site. Results of the $1/N$ expansion for the impurity
critical exponents of uniform and staggered magnetization are
obtained.

\section{LGW criticality}
\label{sec:on}

This section will study the field theory $\mathcal{S}_b^{\bf n} +
\mathcal{S}_{\rm imp}^{\bf n}$ describing an impurity in an
antiferromagnet with an even number of $S=1/2$ spins per unit cell.
As discussed in Section~\ref{sec:intro}, the O(3) non-linear sigma
model formulation in Eqs.~(\ref{sbn}) and (\ref{simpn}) is not
appropriate for our purposes. Instead, we shall use a `soft-spin'
approach which yields a convenient description of the
rotationally-invariant state of the bulk antiferromagnet for $g \geq
g_c$, and of its impurity-induced deformations. The universal
results appear in an expansion in
\begin{equation}
\epsilon = (3-d),
\end{equation}
where $d$ is the spatial dimensionality.

This dimensionality expansions allow us to compute, in principle,
the universal scaling functions, appearing in Eqs.~(\ref{Qscale})
and (\ref{nscale}), which were numerically computed recently in
Ref.~\onlinecite{sandvik3}. The scaling functions clearly depend
upon the geometry of the sample, and the nature of the finite-size
boundary conditions. Such features are not easily captured in a
dimensionality expansion. Consequently the results in this section
are more a ``proof of principle'' that the scaling results apply.
Direct comparison of the results below for scaling functions to the
numerical results are not very useful.

As discussed in Ref.~\onlinecite{qimp1}, the $\epsilon$ expansion is
obtained by replacing the fixed length field ${\bf n}$ by a field
$\bm{\phi}$ whose amplitude is allowed to vary freely. However, we
do not have the freedom to relax the length constraint on the
impurity site because the Berry phase term is only defined for a
unit length field. Consequently, we retain an independent field
${\bf n} (\tau)$ representing the impurity spin, which is now
linearly coupled to $\bm{\phi}$. So we consider the theory
\begin{eqnarray}
\mathcal{Z}_\phi &=& \int \mathcal{D} \bm{\phi} (\vec{x},\tau)
\mathcal{D} {\bf n} (\tau) \delta \left({\bf n}^2 - 1 \right) \exp
\left( - \mathcal{S}_b^\phi -
\mathcal{S}_{\rm imp}^\phi \right) \nonumber \\
\mathcal{S}_{b}^\phi &=& \int d^d x d \tau \left[ \frac{1}{2} \left(
(\partial_\mu \bm{\phi} )^2 + s \bm{\phi}^2 \right) + \frac{g_0}{4!}
\left( \bm{\phi}^2 \right)^2 \right] \nonumber \\
\mathcal{S}_{\rm imp}^\phi &=& i S \int d\tau {\bf A} [ {\bf n}
(\tau)] \cdot \frac{d {\bf n} (\tau)}{d \tau} - \gamma_0 S {\bf n}
(\tau) \cdot \bm{\phi} (0,\tau) \label{zlgw}
\end{eqnarray}
Here $s \sim g$ is the coupling that tunes the system across the
bulk quantum phase transition, and $g_0$ and $\gamma_0$ are the
couplings which were shown in Ref.~\onlinecite{qimp1} to approach
fixed point values in the vicinity of the quantum critical point. In
the $(3-d)$ expansion, these fixed point values are small with $g_0
\sim \gamma_0^2 \sim \epsilon$. It was argued in
Ref.~\onlinecite{qimp2} that this fixed point is identical to that
obtained from the O(3) non-linear sigma model theory appearing in
Eqs.~(\ref{sbn},\ref{simpn}).

We will be interested here in the $s \geq s_c$ regime of
$\mathcal{Z}_\phi$ here, where $\langle \bm{\phi} \rangle = 0$ and
full rotational symmetry is preserved in the absence of the
impurity. As discussed in Section~\ref{sec:intro}, we need to
project on to the state with total $S_z = 1/2$ in the presence of
the impurity. This is easily done here by choosing the following
parameterization for the impurity degree of freedom ${\bf n} (\tau)$
in terms of a complex scalar $\psi (\tau)$:
\begin{equation}
{\bf n} = \left( \frac{\psi+\psi^{\ast}}{2} \sqrt{2 - |\psi |^2},
\frac{\psi- \psi^{\ast}}{2 i} \sqrt{2 - |\psi|^2}, 1 - |\psi|^2
\right). \label{a2}
\end{equation}
The advantage of the representation (\ref{a2}) is that with the
gauge choice
\begin{equation}
{\bf A} ({\bf n}) = \frac{1}{1 + n_z}\left(- n_y, n_x, 0 \right),
\label{a3}
\end{equation}
the Berry phase takes the following form
\begin{equation}
i {\bf A} ({\bf n}) \cdot \frac{d {\bf n}}{d \tau} = \frac{1}{2}
\left( \psi^{\ast} \frac{\partial\psi}{\partial \tau}- \psi
\frac{\partial\psi^{\ast}}{\partial \tau}\right), \label{a4}
\end{equation}
Furthermore, the measure term in the functional integral also has
the simple form
\begin{equation}
\int \mathcal{D} {\bf n} \delta\left( {\bf n}^2 - 1 \right) =  \int
\mathcal{D} \psi \mathcal{D} \psi^{\ast} \label{a5}
\end{equation}
Now, an expansion of the correlators of $\mathcal{Z}_\phi$, in a
functional integral over $\bm{\phi}$ and $\psi$ about the saddle
point with $\bm{\phi} = 0$ and $\psi = 0$, in powers of the
couplings $\gamma_0$ and $g_0$, automatically projects onto the
state with total spin projection $S_z = 1/2$. This is easily
established by applying a uniform magnetic field, and verifying by
the methods of Ref.~\onlinecite{qimp1,qimp2} that the total
magnetization is quantized by a Ward identity associated with the
conservation of spin.

We can now use the above perturbative expansion, using methods
explained at length elsewhere\cite{qimp1,qimp2}, to compute the
expectation values of the magnetization density $\langle Q_z
(\vec{x}) \rangle$ and the N\'eel order parameter $\langle \phi_z
(\vec{x} ) \rangle$. We perform this computation on a sample with
periodic boundary conditions and length $L$ in each spatial
dimension, {\em i.e.\/} a torus $T^d$. The main effect of the finite
boundary conditions is that the momenta $\vec{p}$ are discrete, and
each momentum component is quantized in integer multiples of
$2\pi/L$. The results below are easily generalized to other finite
size geometries and boundary conditions. To leading order in
$\epsilon$, the results are
\begin{eqnarray}
\langle Q_z (\vec{x}) \rangle &=& S \delta^d (\vec{x}) - \gamma_0^2
S \delta^d (\vec{x}) \int \frac{d \omega}{2 \pi} \frac{1}{(i \omega
+ \varepsilon)^2}G(\omega, 0) + 2 \gamma_0^2 S \int \frac{d
\omega}{2 \pi} G(\omega,
\vec{x}) G(\omega, -\vec{x}) \nonumber \\
\langle \phi_z (\vec{x}) \rangle &=& \gamma_0 S G(0, \vec{x}) \left[
1 - \gamma_0^2  \int \frac{d \omega}{2 \pi} \frac{1}{(i \omega +
\varepsilon)^2}G(\omega, 0) \right] \label{d1}
\end{eqnarray}
where $\varepsilon$ is a positive infinitesimal proportional to an
applied magnetic field which selects the $S_z = 1/2$ state. We may
set $\varepsilon=0$ after the frequency integrals have been
performed. The Green's function of the $\bm{\phi}$ field is
\begin{equation}
G(\omega,\vec{x}) = \frac{1}{L^d} \sum_{\vec{p}} \frac{e^{i \vec{p}
\cdot \vec{x}}}{ \omega^2 + \vec{p}^2 + \Delta^2}, \label{g1}
\end{equation}
where $\Delta$ is the spin gap of the bulk antiferromagnet in the
absence of the impurity. Other boundary conditions will only change
the form of $G$. It is easy to check that the spatial integral of
$\langle Q_z \rangle$ is quantized at $S$.

To leading order in $\epsilon$, it would appear that we can set
$\Delta$ equal to the spin gap in the infinite bulk antiferromagnet,
and in particular, set $\Delta =0$ at the critical point $s=s_c$.
However, we will see below that for the particular boundary
conditions we are using here, there are infrared divergencies at
$\Delta=0$ in the expressions for the impurity-induced spin
textures. In such a situation we have to examine the finite $L$
corrections to the value of $\Delta$ at $s=s_c$, which yield a
non-zero $\Delta$ even at the bulk quantum critical point. The value
of $\Delta$ can be computed as described elsewhere\cite{sscrit}, and
to leading order in $\epsilon$, the equation determining $\Delta$ at
the quantum critical point $s=s_c$ is
\begin{equation}
\Delta^2 = \frac{5 g_0}{6} \frac{1}{L^d} \sum_{\vec{p}} \int
\frac{d\omega}{2 \pi} \frac{1}{ \omega^2 + \vec{p}^2 + \Delta^2}
\end{equation}
To leading order in $\epsilon$, only the $\vec{p}=0$ term on the
right-hand-side has to be included; setting $g_0$ equal to its fixed
point value\cite{sscrit} we find for small $\epsilon$
\begin{equation}
\Delta = \left( \frac{20 \pi^2 \epsilon}{11} \right)^{1/3}
\frac{1}{L}.
\end{equation}
Note that $L\Delta $ is a universal number at $s=s_c$, which is the
main result we will need below to establish the universality of the
spin texture.

Returning to the expressions in Eq.~(\ref{d1}), we now want to
manipulate them into the forms of Eq.~(\ref{Qscale}) and
(\ref{nscale}). However, the presence of the $\delta^d (\vec{x})$ in
Eq.~(\ref{d1}) makes the $\vec{x}$ dependence singular. These
singularities are in fact an artifact of the present perturbative
expansion in real space, and are not expected to be present once the
expansion is resummed. This is evident by examining the results in
momentum space, where the results are a smooth function of momentum.
In this manner we obtain after applying Eq.~(\ref{g1}) to
Eq.~(\ref{d1})
\begin{eqnarray}
\langle Q_z (\vec{p}) \rangle &=& S \left[ 1 -
\frac{\gamma_0^2}{L^d} \sum_{\vec{q}} \frac{1}{2E_{\vec{q}}} \left(
\frac{1}{E_{\vec{q}}^2} - \frac{2}{E_{\vec{p}+\vec{q}} (
E_{\vec{p}+\vec{q}}+ E_{\vec{q}})} \right)
 \right]\nonumber \\
\langle \phi_z (\vec{p} ) \rangle &=& \frac{\gamma_0 S}{\vec{p}^2 +
\Delta^2} \left[ 1 - \frac{\gamma_0^2}{L^d} \sum_{\vec{q}}
\frac{1}{2E_{\vec{q}}^3}
 \right]\label{d2}
\end{eqnarray}
where $E_{\vec{p}} = \sqrt{ \vec{p}^2 + \Delta^2}$. Now
Eqs.~(\ref{d2}) can be evaluated at the fixed point value of
$\gamma_0$, and to leading order in $\epsilon$ they are seen to
yield results consistent with the following scaling forms which can
be deduced from Eqs.~(\ref{Qscale},\ref{nscale})
\begin{eqnarray}
\langle Q_z (\vec{p}) \rangle &=& \widetilde{\Phi}_Q ( \vec{p} L)
\nonumber \\
\langle \phi_z (\vec{p}) \rangle &=& L^{(d+1-\eta)/2}
\widetilde{\Phi}_n ( \vec{p} L)
\end{eqnarray}
The explicit results for the scaling functions to leading order in
$\epsilon$ are
\begin{eqnarray}
\widetilde{\Phi}_Q (\vec{y}) &=& S \left[ 1 - 2 \pi^2 \epsilon
\sum_{\vec{x}}  \frac{1}{2\mathcal{E}_{\vec{x}}} \left(
\frac{1}{\mathcal{E}_{\vec{x}}^2} -
\frac{2}{\mathcal{E}_{\vec{y}+\vec{x}} (
\mathcal{E}_{\vec{y}+\vec{x}}+ \mathcal{E}_{\vec{x}})} \right)
 \right]\nonumber \\
\widetilde{\Phi}_n (\vec{y}) &=& \frac{\pi S \sqrt{2
\epsilon}}{\vec{y}^2 + L^2\Delta^2} \left[ 1 - 2 \pi^2 \epsilon
\left( \mbox{a finite number} \right) \right] \label{d3}
\end{eqnarray}
where now $\vec{x}$ and $\vec{y}$ are {\em three\/} dimensional
momenta whose components are quantized in integer multiples of $2
\pi$ (except in the integral in the second equation), and
$\mathcal{E}_{\vec{x}} = \sqrt{\vec{x}^2 + L^2 \Delta^2}$. It is
easily checked that these expressions are free of infrared and
ultraviolet divergencies, and so yield universal results because $L
\Delta$ is a universal number.

From the above expression, we observe that $\widetilde{\Phi}_{Q}
(|\vec{y}| \rightarrow \infty) = S (1 - (\epsilon/2) \ln
|\vec{y}|)$, which we assume exponentiates to $\widetilde{\Phi}_{Q}
(|\vec{y}| \rightarrow \infty) \sim |\vec{y}|^{-\epsilon/2}$. From
the short distance behavior of the spin texture discussed in
Ref.~\onlinecite{sandvik3}, we expect that $\widetilde{\Phi}_{Q}
(|\vec{y}| \rightarrow \infty) \sim |\vec{y}|^{-\eta'/2}$, where
$\eta'$ is the scaling dimension of the boundary spin\cite{qimp1}.
So we obtain the value $\eta' = \epsilon$, which is consistent with
earlier results\cite{qimp1}. Similarly, from the short distance
behavior discussed in Ref.~\onlinecite{sandvik3}, we also have
$\widetilde{\Phi}_{n} (|\vec{y}| \rightarrow \infty) \sim
|\vec{y}|^{-2 + (\epsilon+\eta-\eta')/2}$. So with $\eta \sim
\mathcal{O}(\epsilon^2)$ and $\eta' = \epsilon$, we have
$\widetilde{\Phi}_{n} (|\vec{y}| \rightarrow \infty) \sim
|\vec{y}|^{-2}$, which is consistent with Eq.~(\ref{d3}).

\section{Deconfined criticality}
\label{sec:cpn}

This section describes the N\'eel-VBS transition in square lattice
quantum antiferromagnets with a single $S=1/2$ per unit cell. As
discussed in Section~\ref{sec:intro}, the response of a non-magnetic
impurity is described by the action $\mathcal{S}_b^z +
\mathcal{S}_{\rm imp}^z$ in Eqs.~(\ref{sbz},\ref{simpz}) for a
complex SU($N$) spinon field $z_\alpha$ and a non-compact U(1) gauge
field $A_\mu$. Here we will describe the $1/N$ expansion of its
universal critical properties. Note that in what follows we have
rescaled the spinon field $z$, to remove the coupling constant $g$
from the action (\ref{sbz}), in favour  of a rescaled constraint
$z^{\dagger}_\alpha z_\alpha = 1/g$. This constraint is enforced
with a local Lagrange multiplier $\lambda$, so that the bulk action
becomes,
\begin{equation}
\mathcal{S}_{b}^z =  \int d\tau \int d^2 x \left[ |(\partial_\mu -i
A_\mu )z_\alpha|^2 + i \lambda (|z_{\alpha}|^2 - \frac{1}{g}) +
\frac{1}{2 e^2} (\epsilon_{\mu\nu\lambda}
\partial_\nu A_\lambda)^2 \right]\,. \label{sbzr}
\end{equation}

It is useful to define SU($N$) generalizations of the SU(2)
observables introduced in Section~\ref{sec:intro}. The uniform
magnetization density ${\bf Q}$ generalizes to $Q^a$, which is  the
temporal component of a current associated with the SU($N$) rotation
symmetry,
 \begin{equation} Q^a = z^{\dagger} T^a D_\tau z - (D_\tau z)^{\dagger} T^a z\end{equation}
 (where $D_\mu = \partial_\mu - i A_\mu$ is the covariant derivative)
 while the N\'eel order ${\bf n}$ in Eq.~(\ref{nzz}) becomes
 the staggered magnetization operator
 \begin{equation} n^a = z^{\dagger} T^a z\end{equation}
where $T^a$ are generators of the SU($N$) algebra. We will describe
the spatial dependence of the expectation values of these operators
for two cases: a finite system of size $L$ at the critical point
$g=g_c$ in Section~\ref{sec:finite}, and the
 infinite system in the N\'eel phase with broken SU($N$) symmetry in
 Section~\ref{sec:neel}.

\subsection{Critical point in a finite system}
\label{sec:finite}

We tune the system to the critical point $g = g_c$ of the infinite
volume zero temperature model, and then consider the system on a
spatial torus of length $L$. We use periodic boundary conditions for
all fields.\footnote{In principle, on a spatial torus, we can
certainly have a finite magnetic ($F_{ij}$) flux, which would
correspond to non-periodic boundary conditions. However, finite flux
sectors are expected to be separated from vacuum by an energy gap,
and hence are suppressed at $T = 0$.} As we discussed in
Section~\ref{sec:intro}, the ground state in the absence of an
impurity is a spin-singlet, while adding an impurity yields a ground
state which transforms under the fundamental representation of
SU($N$). This ground state has a single spinon in it, and we argued
that the projection onto this state can be performed by
Eq.~(\ref{matel}). For an additional test of our projection
formalism, see the appendix, where we compute the $U(1)$ (electric)
charge density in the presence of the impurity.

Before we address the explicit computation of (\ref{matel}), we
discuss scaling forms that our results should obey.

\subsubsection{Scaling Forms}
\label{sec:scaling}

We are interested in computing the uniform and staggered
magnetization densities. Recall, that since the uniform
magnetization is a zeroth component of a conserved current, it
receives no renormalizations. Therefore, utilizing the SU($N$)
symmetry, we have the general scaling form,
\begin{equation} \langle
\alpha| Q^a(\vec{x})|\beta\rangle = \frac{1}{L^2}
\Phi_Q\Big(\frac{\vec{x}}{L}\Big)\,T^a_{\alpha \beta}.
\end{equation}
The leading $1/L^2$ prefactor corresponds to the scaling dimension
$\Delta_Q=d=2$ of the magnetization density. Moreover, by
conservation of total SU($N$) charge,
\begin{equation}
\label{integconst} \int d^2 r \,\Phi_Q(\vec{r}) = -1
\end{equation}
where the integral is over $0 < r_1, r_2 < 1$. Similarly, for the
case of the staggered magnetization,
\begin{equation}
\langle \alpha|n^a(\vec{x})|\beta\rangle =
\Lambda^{\eta_n}\left(\frac{1}{L}\right)^{1-\eta_n}
\Phi_n\left(\frac{\vec{x}}{L}\right) T^a_{\alpha \beta}
\end{equation}
Here $\eta_n$ is the anomalous dimension of the staggered
magnetization operator $n^a(x)$, $\Delta_n = dim[n^a] = 1 - \eta_n$.
This exponent is related to the exponent $\eta$ in
Eq.~(\ref{nscale}), and their values were computed previously
\cite{irkhin} in the $1/N$ expansion for arbitrary spacetime
dimension $2 < D < 4$:
\begin{equation}
\eta_n = \frac{1}{2} (D-2-\eta) = \frac{1}{N} \frac{16
\Gamma(D-2)}{\Gamma(2-D/2) \Gamma(D/2-1)^3} + O(1/N^2)
\stackrel{D=3}{=}\frac{16}{\pi^2 N} + O(1/N^2)\,.
\end{equation}
The function $\Phi_Q$ is completely universal, whereas $\Phi_n$ is
universal only up to an overall scale. In particular, $\Phi_n$ does
not have any property analogous to (\ref{integconst}).

Of particular interest is the behavior of the functions
$\Phi_Q(\vec{r})$, $\Phi_n(\vec{r})$ for $\vec{r} \to 0$. We make a
hypothesis that $n^a(\vec{x},\tau)$ and $Q^a(\vec{x},\tau)$ flow to
the same operator $S^a(\tau)$ as $\vec{x}$ approaches the Wilson
line,
\begin{eqnarray}
\label{OPE} \lim_{|\vec{x}|\to
0}Q^a(\vec{x},\tau) = \frac{c_Q}{|\vec{x}|^{-\Delta^{Q}_{\rm imp}}}S^a(\tau)\\
\nn\lim_{|\vec{x}|\to 0}n^a(\vec{x},\tau) =
\frac{c_n}{|\vec{x}|^{-\Delta^{n}_{\rm imp}}}S^a(\tau)
\end{eqnarray}
Calculations in the $\epsilon$ expansion supporting this hypothesis
have been given in Ref.~\onlinecite{ImpurityKSBC}. We have performed
analogous calculations in the $1/N$ expansion again confirming the
OPE (\ref{OPE}). Technically, this impurity OPE program consists of
the following steps. First one considers the (multiplicative)
renormalization of the operator $n^a(\vec{x} = 0)$, by studying its
insertion into the two point function of the $z$ field (this consist
of the usual bulk renormalization, plus an additional
renormalization of the logarithmic divergences that appear as
$\vec{x} \to 0$). Once $n^a(\vec{x} = 0)$ operator is renormalized,
one considers the insertion of $Q^a(\vec{x}\to 0)$ into the two
point function of the $z$ field. The highest divergence as
$|\vec{x}| \to 0$ is power-like, $1/|\vec{x}|$,  modified by
logarithms at higher orders in $1/N$. This leading divergence can be
cancelled by a $n^a(\vec{x}= 0)$ counterterm (with a coefficient
that diverges as $\vec{x} \to 0$). This procedure gives one a way to
construct order by order in $1/N$, the impurity operator $S^a(\tau)$
(which is essentially a regularized $n^a(\vec{x} = 0, \tau)$), and
compute the anomalous dimensions $\Delta^Q_{\rm imp}$,
$\Delta^n_{\rm imp}$ as well as coefficients $c_Q$, $c_n$ (the later
are renormalization scheme dependent). As the computation of the OPE
in the $1/N$ expansion essentially follows that in the $\epsilon$
expansion presented in Ref. \onlinecite{ImpurityKSBC}, we shall not
include it here. We only note that in this way, we have been able to
explicitly check the OPE (\ref{OPE}) to order $1/N^2$, obtaining
$\Delta^n_{\rm imp}$ to order $1/N^2$ and $\Delta^Q_{\rm imp}$ to
order $1/N$ (this is lower order than the corresponding result for
$\Delta^n_{\rm imp}$ as $c_Q/c_n$ is of order $1/N$). Explicit
results in this expansion will appear in Section~\ref{sec:neel}.

Calculations of $\Phi_Q$ and $\Phi_n$ given below provide additional
support for the OPE (\ref{OPE}). Note that the exponents
$\Delta^{Q}_{\rm imp}$ and $\Delta^{n}_{\rm imp}$ are not
independent. Indeed, let the correlator
\begin{equation}
\langle S^a(\tau) S^b(0) \rangle \sim \frac{1}{\tau^{2\, \Delta_S}}
\,\delta_{ab} \,.
\end{equation}
The exponent $\Delta_S$ is related to the boundary spin exponent
$\eta'$ used in Refs.~\onlinecite{qimp1,sandvik3} by $\eta' = 2
\Delta_S$. Then,
\begin{equation}
\Delta_S = \Delta_Q + \Delta^{Q}_{\rm imp} = \Delta_n +
\Delta^n_{\rm imp}
\end{equation}
Recalling, $\Delta_Q = 2$, $\Delta_n = 1-\eta_n$,
\begin{equation}
\label{impdimrel}\Delta^Q_{\rm imp} = \Delta^n_{\rm imp} - 1 -
\eta_n \,.
\end{equation}
Our explicit results for the profiles $\Phi_Q, \Phi_n$ confirm the
relation (\ref{impdimrel}) to leading (zeroth) order in $1/N$, see
below. We have also been able to  check this relation to order $1/N$
using the impurity OPE program summarized above: to this order,
$\Delta^Q_{\rm imp} = -1 - \eta_n$, as $\Delta^n_{\rm imp} \sim
O(1/N^2)$. The result of our evaluation of $\Delta^n_{\rm imp}$ to
$O(1/N^2)$ will appear later in
Eqs.~(\ref{deltanimp}),(\ref{Deltanumer}).

Note that the OPE (\ref{OPE}) is sensitive only to short distance
physics, and, thus, coefficients $c_Q$, $c_n$ should be independent
of the system size $L$ as well as the deviation from the critical
point (all this IR information is, however, contained in the
impurity operator $S^a$). Thus, the ratio,
\begin{equation}
\frac{c_Q}{c_n} = \lim_{|\vec{x}| \to 0} |\vec{x}|^{\Delta^n_{\rm
imp}-\Delta^Q_{\rm imp}} \frac{\langle Q^a(\vec{x}) \rangle}{\langle
n^a(\vec{x})\rangle}= \lim_{|\vec{x}|\to 0}|\vec{x}|^{1 + \eta_n}
\frac{\langle Q^a(\vec{x}) \rangle}{\langle n^a(\vec{x})\rangle}
\end{equation}
although non-universal, should be constant throughout the scaling
regime (once the regularization scheme is chosen). We shall check
this fact below to leading order in $1/N$ by comparing the short
distance behaviour (controlled by the OPE) of uniform and staggered
magnetization densities at the critical point and in the N\'eel
phase.

\subsubsection{Projection onto the Single Spinon State}

Now we return to the evaluation of the matrix elements
(\ref{matel}). Although it is possible to obtain all the results
presented below directly from Eq. (\ref{matel}) it is technically
somewhat simpler to use instead,
\begin{equation} \label{matelnong}
\langle \alpha|O(\vec{x})|\beta\rangle = \lim_{\mathcal{T}\to
\infty}\frac{\langle z_\alpha(\vec{k},\mathcal{T}/2) O(\vec{x},0)
 z^{\dagger}_{\beta}(\vec{k}',-\mathcal{T}/2)\rangle_{\rm imp}}{\langle z_\alpha(\vec{k},\mathcal{T}/2)
 z^{\dagger}_{\alpha}(\vec{k}',-\mathcal{T}/2)\rangle_{\rm imp}}
\end{equation}
Here, $z_\alpha(\vec{k},\tau) = \int d^2 x z_\alpha(\vec{x},\tau)
e^{-i \vec{k} \vec{x}}$ and the subscript ``imp'' indicates that the
correlator should be computed in a theory with the action
$\mathcal{S}_{b}^z + \mathcal{S}_{\rm imp}^z$ which includes the
impurity term. Effectively, we have extended the Wilson line, which
in (\ref{matel}) stretched from the point where a spinon was created
to the point where it was destroyed, to run from $\tau = -\infty$ to
$\tau = \infty$. In addition, we have taken our ``incoming" and
``outgoing" spinon to be in momenta $\vec{k}$ and $\vec{k}'$ states.
This makes the numerator and denominator of (\ref{matelnong})
non-gauge invariant. Nevertheless, we expect that this non-gauge
invariance comes solely from the matrix element for creating the
ground state of the system by acting on the vacuum with
$z^{\dagger}$ and cancels out between the numerator and denominator
of (\ref{matelnong}).

Since the impurity term Eq.~(\ref{simpz}) breaks spatial (but not
temporal) translational invariance, for $\mathcal{T} \to \infty$ we
expect to obtain the ground state irrespective of which $\vec{k},
\vec{k}'$ we started with. Nevertheless, it will be most convenient
in our perturbative treatment to work with $\vec{k} = \vec{k}' = 0$.

Since the external charge does not break SU($N$) symmetry and time
translation symmetry, we have,
\begin{equation}
\langle z_{\alpha}
(x) z^{\dagger}_{\beta} (x')\rangle_{\rm imp} = \delta_{\alpha
\beta} D(\vec{x},\vec{x}',\tau - \tau')
\end{equation}
We let,
\begin{equation}
D(\vec{x},\vec{x}',\tau) = \frac{1}{L^2}
\sum_{\vec{p},\vec{p}'} \int \frac{d \omega} {2 \pi}
D(\vec{p},\vec{p}',\omega) e^{i \vec{p} \vec{x}} e^{-i \vec{p}'
\vec{x}'} e^{i \omega \tau}
\end{equation}
We write,
\begin{equation} \langle z_{\alpha}(y) O(x)
z^{\dagger}_{\beta}(y')\rangle_{\rm imp} = \int dv dv' D(y,v)
O_{\alpha \beta} (v, x, v') D(v', y')
\end{equation}
Fourier
transforming,
\begin{equation}
O_{\alpha \beta}(v,x,v') = \frac{1}{L^2}\sum_{\vec{p}}
\frac{1}{L^2}\sum_{\vec{p}'}\! \int \frac{d \omega}{2 \pi}\! \int
\frac{d \omega'}{2 \pi} O_{\alpha
\beta}(\vec{p},\vec{q},\vec{p}',\omega,\omega') e^{i \vec{p}
\vec{v}} e^{-i \vec{p}' \vec{v}'} e^{i \vec{q} \vec{x}} e^{i \omega
v_{\tau}} e^{- i \omega' {v_{\tau}}'} e^{i (\omega' - \omega)
x_{\tau}}
\end{equation}
where we use the notation that the
three-vector $x$ has spatial components $\vec{x}$ and temporal
component $x_\tau$. So,
\begin{eqnarray}
&&\langle
z_\alpha(\vec{k},\mathcal{T}/2) O(\vec{x},0)
 z^{\dagger}_{\beta}(\vec{k}',-\mathcal{T}/2)\rangle_{\rm imp} =\nn\\
 &&\sum_{\vec{p},\vec{p}',\vec{q}} \int \frac{d\omega}{2 \pi} \frac{d \omega'}{2
 \pi} D(\vec{k}, \vec{p}, \omega) O_{\alpha \beta}(\vec{p},
 \vec{q}, \vec{p}',\omega, \omega') D(\vec{p}', \vec{k}',\omega')
 e^{i \omega \mathcal{T}/2} e^{i \omega' \mathcal{T}/2} e^{i \vec{q} \vec{x}}
\end{eqnarray}
As we perform the integral over $\omega$, $\omega'$, we pick up
poles of the propagators $D$ in the $\Im(\omega) > 0$, $\Im(\omega')
>0$ planes (we expect that $O_{\alpha \beta}$ is analytic in
$\omega$). In the limit $\mathcal{T} \to \infty$ only the
contribution from the pole with smallest imaginary part survives.
Let this pole be at $\omega = i m$ and denote by ${\rm
Res}(\vec{k},\vec{p})$ the residue of $D(\vec{k},\vec{p}, \omega)$
at this pole. Then,
\begin{equation}
\langle\! z_\alpha\!(\vec{k},\mathcal{T}/2) O(\vec{x},0)
 z^{\dagger}_{\beta}(\vec{k}',-\mathcal{T}/2)\!\rangle_{\rm imp}\!
 \to\!\!\!
\sum_{\vec{p},\vec{p}',\vec{q}}\! (i {\rm Res}(\vec{k},\vec{p}))(i
{\rm Res}(\vec{p}',\vec{k}')) O_{\alpha \beta}(\vec{p},
 \vec{q}, \vec{p}',i m, i m)  e^{i \vec{q} \vec{x}} e^{- m \mathcal{T}}
 \end{equation}
Similarly, the denominator of (\ref{matelnong}) is,
\begin{equation}
\langle z_{\alpha}(\vec{k},\mathcal{T}/2)
z^{\dagger}_{\alpha}(\vec{k}',-\mathcal{T}/2)\rangle_{\rm imp} \to
L^2 i {\rm Res}(k, k') e^{-m \mathcal{T}}
\end{equation}
Finally,
\begin{equation}
\label{matres} \langle \alpha|
O(\vec{x})|\beta\rangle = \frac{1}{L^2} \sum_q \langle
\alpha|O(\vec{q})|\beta\rangle \,e^{i \vec{q} \vec{x}}
\end{equation}
with,
\begin{equation}
\label{matres1} \langle \alpha| O(\vec{q})|\beta\rangle =
\sum_{\vec{p},\vec{p}'} \frac{(i {\rm Res}(\vec{k},\vec{p}))(i {\rm
Res}(\vec{p}',\vec{k}'))}{i {\rm Res}(k,k')} O_{\alpha
\beta}(\vec{p},
 \vec{q}, \vec{p}',i m, i m)
 \end{equation}

\subsubsection{Large $N$ expansion of $\mathbb{CP}^{N-1}$ theory in finite
volume}

We now compute the expression (\ref{matres1}) using the large $N$
expansion in finite volume. First, consider the $N = \infty$ limit.
The gap equation reads,
\begin{equation} \frac{1}{L^2} \sum_{\vec{p}}
\int \frac{d \omega}{2 \pi} \frac{1}{\omega^2 + \vec{p}^2 + m_0^2} =
\frac{1}{g N}
\end{equation}
and to this order in $N$, $m_0^2 = i
\langle \lambda \rangle$. In the infinite volume, the critical
coupling $g = g_c$ is obtained when the gap $m_0$ vanishes,
\begin{equation}
\label{gapeqcr} \frac{1}{g_c N} = \int \frac{d^3
p}{(2 \pi)^3} \frac{1}{p^2}
\end{equation}
However, once we make the
spatial volume finite, a non-zero $m_0$ is
generated even at the critical point. %(this is the spatial equivalent
%of the thermal mass).
Thus, setting $g = g_c$, using Eq. (\ref{gapeqcr}) and poisson
resumming, we obtain,
\begin{equation}
\sum_{\vec{n} \in {\bf Z}^2 } \int \frac{d \omega}{2 \pi} \int
\frac{d^2 p}{(2 \pi)^2} e^{ i \vec{p} \,\vec{n} L} \frac{1}{\omega^2
+ \vec{p}^2 + m_0^2} = \int \frac{d \omega}{2 \pi} \int \frac{d^2
p}{(2 \pi)^2} \frac{1}{\omega^2 + \vec{p}^2}
\end{equation}
On the left-hand side, only the $\vec{n} = 0$ term diverges in
the UV. However, this divergence cancels with the divergence of the
right-hand side. Thus, performing all integrals,
\begin{equation}
\label{gapL} \sum_{\vec{n} \neq 0} \frac{1}{4 \pi |\vec{n}|} e^{-
m_0 |\vec{n}| L} = \frac{m_0 L}{4 \pi}
\end{equation} The solution of the Eq.~(\ref{gapL}) is,
\begin{equation}
m_0 = \theta \frac{1}{L}
\end{equation}
where $\theta$ is a constant that can be obtained by
solving (\ref{gapL}) numerically to be, $\theta \approx 1.51196$.

Thus, at leading order the propagator,
\begin{equation}
D_0(\vec{k},\vec{k}',\omega) = \delta_{\vec{k},\vec{k}'}
\frac{1}{\omega^2 + \vec{k}^2 + m^2_0}
\end{equation}
and the lowest
pole is at $\vec{k} = 0$, $\omega = i m_0$ and, $i {\rm
Res}(\vec{k},\vec{p}) = \delta_{\vec{k},0} \delta_{\vec{p},0}
\frac{1}{2 m_0}$.

To develop the $1/N$ expansion, we will need to find the $A_{\mu}$
and $\lambda$ propagators. The dynamically generated self-energy for
$A_{\mu}$ is to leading order,
\begin{equation} K_{\mu \nu}(p) = - N
\frac{1}{L^2} \sum_{\vec{q}} \int \frac{d q_{\tau}}{2 \pi}
\left(\frac{(2 q - p)_{\mu} (2 q-p)_{\nu}}{((q-p)^2 + m^2_0)(q^2 +
m^2_0)} - \frac{2 \delta_{\mu
\nu}}{(q^2+m_0^2)}\right)
\end{equation}
This self energy is always more singular near the critical point
than the bare Maxwell term in $\mathcal{S}_b^z$, and so we will work
with $e^2 = \infty$ for the rest of this paper. To find the photon
propagator, ${\cal D}_{\mu \nu}(p)$, we also need to fix a gauge.
Practically, for the calculations to follow, we will only need the
static electromagnetic propagator ${\cal
D}_{\tau\tau}(\vec{p},p_{\tau} = 0) = K_{\tau\tau}(\vec{p},p_{\tau}
= 0)^{-1}$, which is a gauge invariant quantity. We also note that
in the infinite volume limit,
\begin{eqnarray}
K_{\mu \nu}(q) &=& K(q) (q^2 \delta_{\mu \nu} - q_{\mu}
q_{\nu})\\K(q) &=& N A\, q^{D-4} \label{Kc}\end{eqnarray} where the
constant $A$ is given by,
\begin{equation}
A = \frac{1}{(4
\pi)^{D/2}}\frac{(D-2) \Gamma(2-D/2)
\Gamma(D/2-1)^2}{\Gamma(D)}
\end{equation} Here $D$ is the space-time
dimension. In our case, $D = 3$ and $A = \frac{1}{16}$.

Likewise, the self-energy for $\lambda$ is to leading order,
\begin{equation}
\Pi(p) = N \frac{1}{L^2} \sum_{\vec{q}} \int \frac{d q_{\tau}}{2
\pi} \frac{1}{(q^2 + m^2_0)((q-p)^2 + m^2_0)} \label{Pic}
\end{equation}
In the infinite volume limit,
\begin{equation}
\Pi(p) = N B p^{D-4}
\end{equation}
where the constant $B$ is given
by,
\begin{equation} B =\frac{1}{(4
\pi)^{D/2}}\frac{\Gamma(2-D/2)\Gamma(D/2-1)^2}{\Gamma(D-2)}\end{equation}
For $D = 3$, $B = \frac{1}{8}$.

\subsubsection{Matrix Elements}

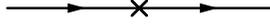
\begin{figure}[t]
\begin{fmffile}{gMNlead}
\fmfset{arrow_len}{2.5mm}
\begin{fmfgraph*}(100,60)
\fmfleft{l} \fmfright{r} \fmf{plain_arrow}{l,v1}
\fmf{plain_arrow}{v1,r} \fmfv{d.sh=cross,d.f=empty,d.si=8thin}{v1}
\end{fmfgraph*}
\end{fmffile}
\caption{The insertion of $Q^a$ into the $z$
propagator.}\label{diaMzlead}
\end{figure}

Now, let us compute the matrix elements of operator $Q^a(x)$. The
insertion of $Q^a$ into the $z$ propagator, to leading order in
$1/N$ is given by diagram in Fig. \ref{diaMzlead}, so
\begin{equation}
Q^a_{\alpha\beta}(\vec{p},\vec{q},\vec{p}',\omega,\omega') = i
(\omega + \omega') T^a_{\alpha \beta}
\delta_{\vec{q},\vec{p}'-\vec{p}}
\end{equation}
So utilizing formula (\ref{matres1}), with $\vec{k} = \vec{k}' = 0$,
we obtain,
\begin{equation}
\label{Maq0} \langle \alpha|Q^a(\vec{q})|\beta\rangle  =
-T^a_{\alpha \beta} \delta_{\vec{q} 0}
\end{equation}
{\em i.e.\/},
\begin{equation}
\langle \alpha|Q^a(\vec{x})|\beta\rangle = -\frac{1}{L^2}
T^a_{\alpha \beta}
\end{equation}
and the function $\Phi_Q(\vec{r}) = -1$, satisfies the normalization
condition (\ref{integconst}). So at leading order in the $1/N$
expansion the magnetization in the presence of an impurity is
spatially uniform. The system with the impurity simply consists of a
free spinon in the zero momentum state. The effects of the
interaction with the impurity appear only at next order in $1/N$.

Similarly, for the staggered magnetization, the insertion of
$n^a(x)$ into the $z$ propagator, to leading order is given by the
same diagram in Fig. \ref{diaMzlead}, except the cross now stands
for $n^a$.

\begin{equation}
n^a_{\alpha \beta}(\vec{p},\vec{q},\vec{p}',\omega,\omega')=
\delta_{\vec{q},\vec{p}'-\vec{p}} T^a_{\alpha \beta}
\end{equation}
so that,
\begin{equation}
\label{Naqlead} \langle \alpha | n^a(\vec{q})|\beta\rangle =
\frac{1}{2 m_0} \delta_{\vec{q}0} T^a_{\alpha \beta}
\end{equation}
and,
\begin{equation}
\langle\alpha|n^a(\vec{x})|\beta\rangle = \frac{1}{2 \theta L}
T^a_{\alpha \beta}
\end{equation}
So the staggered magnetization at leading order in $1/N$ is also
uniform, $\Phi_n(\vec{r}) = \frac{1}{2 \theta}$.

Now, let's include the $1/N$ corrections.

We will concentrate on corrections to $\langle \alpha| O(\vec{q})
|\beta \rangle$, for $O = Q^a, n^a$, with $\vec{q} \neq 0$ (where
the leading O(1) term vanishes). These turn out to be much simpler
to compute than corrections for $\vec{q} = 0$. Moreover, for $Q^a$,
we know by SU($N$) charge conservation that the $N = \infty$ result
(\ref{Maq0}) at $\vec{q} = 0$ receives no further corrections. Thus,
to order $1/N$,
\begin{eqnarray}
\label{Ocor} \langle \alpha | O(\vec{q})|\beta \rangle
&\stackrel{\vec{q} \neq 0}{=}& i {\rm Res}(0,-\vec{q})_1 \,O_{\alpha
\beta}(-\vec{q}, \vec{q}, 0, i m_0, i m_0)_0 + i {\rm
Res}(\vec{q},0)_1 \,O_{\alpha \beta} (0,\vec{q},\vec{q},i m_0, i
m_0)_0\nn\\ &+& i {\rm Res}(0,0)_0 \, O_{\alpha \beta}(0,\vec{q},0,
i m_0, i m_0)_1
\end{eqnarray}
where the subscripts
$0$, $1$ indicate the order in $1/N$ to which the quantity has to be
computed.

\begin{figure}[t]
\begin{fmffile}{graphsprop}
\fmfset{arrow_len}{2.5mm}
\begin{fmfgraph*}(100,60)
\fmfleft{l} \fmfright{r} \fmf{plain_arrow}{l,v1}
\fmf{plain_arrow}{v2,r} \fmf{plain_arrow}{v1,v2} \fmffreeze
\fmf{wiggly,left}{v1,v2}
\end{fmfgraph*}
\hspace{1cm}
\begin{fmfgraph*}(100,60)
\fmfleft{l} \fmfright{r} \fmf{plain_arrow}{l,v1}
\fmf{plain_arrow}{v2,r} \fmf{dashes,left}{v1,v2}\fmffreeze
\fmf{plain_arrow}{v1,v2}
\end{fmfgraph*}
\hspace{1cm}
\begin{fmfgraph*}(100,60)
\fmfleft{l} \fmfright{r} \fmftop{t} \fmf{plain_arrow}{l,v1}
\fmf{plain_arrow}{v1,r} \fmffreeze \fmf{wiggly}{v1,t}
%\fmfv{d.sh=cross,d.f=empty,d.si=8thin}{v1}
\fmfv{d.sh=square,d.f=full,d.si=6thin}{t}
\end{fmfgraph*}
\end{fmffile}
\caption{$1/N$ corrections to $z$ self-energy.}\label{diazself}
\end{figure}
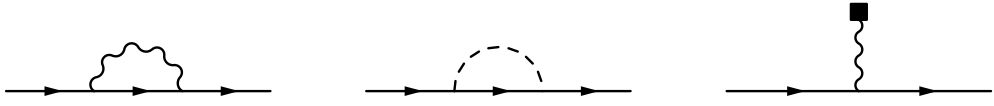

The $1/N$ corrections to the $z$ self-energy are shown in Fig.
\ref{diazself} (we drop $\lambda$ tadpole diagrams).
Of these only the last one couples to the impurity and, therefore,
breaks translational invariance. So, letting,
\begin{equation}
D(\vec{k},\vec{k}',\omega) =
D_0(\vec{k},\vec{k}',\omega) - \sum_{\vec{p},\vec{p}'}
D_0(\vec{k},\vec{p},\omega) \Sigma(\vec{p},\vec{p}',\omega)
D(\vec{p}',\vec{k}',\omega)
\end{equation}
\begin{equation}
\Sigma(\vec{k},\vec{k}',\omega) \stackrel{\vec{k} \neq \vec{k}'}{=}
\frac{1}{L^2} 2 i \omega {\cal D}_{\tau\tau}(\vec{k}-\vec{k}',0) +
O(1/N^2)
\end{equation}
and
\begin{equation}
D(\vec{k},\vec{k'},\omega) \stackrel{\vec{k} \neq \vec{k}'}{=} -
\frac{1}{L^2} 2 i \omega {\cal D}_{\tau\tau}(\vec{k}-\vec{k}',0)
\frac{1}{\omega^2 + \vec{k}^2 + m^2_0}\frac{1}{\omega^2+\vec{k}'^2 +
m^2_0} + O(1/N^2)
\end{equation}
So the residue,
\begin{equation}
\label{Rescor} i {\rm Res}(0, -\vec{q}) = i {\rm Res} (\vec{q},0)
\stackrel{\vec{q} \neq 0}{=} \frac{1}{L^2} \frac{1}{\vec{q}^2}{\cal
D}_{\tau\tau}(\vec{q},0) + O(1/N^2)
\end{equation}
Note that at this
order renormalization of the location of the pole $\omega = i m
\stackrel{N=\infty}{=} i m_0$ can be neglected.

\begin{figure}[h]
\begin{fmffile}{graphsM}
\fmfset{arrow_len}{2.5mm}
\begin{fmfgraph*}(100,60)
\fmfleft{l} \fmfright{r} \fmf{plain_arrow}{l,v1}
\fmf{plain_arrow}{v2,r} \fmf{plain_arrow}{v1,v3}
\fmf{plain_arrow}{v3,v2} \fmf{plain_arrow}{v2,r}
\fmfv{d.sh=cross,d.f=empty,d.si=8thin}{v3} \fmffreeze
\fmf{wiggly,left}{v1,v2}
\end{fmfgraph*}
\hspace{1cm}
\begin{fmfgraph*}(100,60)
\fmfleft{l} \fmfright{r} \fmf{plain_arrow}{l,v1}
\fmf{plain_arrow}{v2,r} \fmf{plain_arrow}{v1,v3}
\fmf{plain_arrow}{v3,v2} \fmf{plain_arrow}{v2,r}
\fmfv{d.sh=cross,d.f=empty,d.si=8thin}{v3} \fmffreeze
\fmf{dashes,left}{v1,v2}
\end{fmfgraph*}
\hspace{1cm}
\begin{fmfgraph*}(100,60)
\fmfleft{l} \fmfright{r} \fmf{plain_arrow}{l,v1}
\fmf{plain_arrow}{v2,r} \fmf{plain_arrow}{v1,v2} \fmffreeze
\fmf{wiggly,left}{v1,v2} \fmfv{d.sh=cross,d.f=empty,d.si=8thin}{v2}
\end{fmfgraph*}
\\
\begin{fmfgraph*}(100,60)
\fmfleft{l} \fmfright{r} \fmf{plain_arrow}{l,v1}
\fmf{plain_arrow}{v2,r} \fmf{plain_arrow}{v1,v2} \fmffreeze
\fmf{wiggly,left}{v1,v2} \fmfv{d.sh=cross,d.f=empty,d.si=8thin}{v1}
\end{fmfgraph*}
\hspace{1cm}
\begin{fmfgraph*}(100,60)
\fmfleft{l} \fmfright{r} \fmftop{t} \fmf{plain_arrow}{l,v1}
\fmf{plain_arrow}{v1,r} \fmffreeze \fmf{wiggly}{v1,t}
\fmfv{d.sh=cross,d.f=empty,d.si=8thin}{v1}
\fmfv{d.sh=square,d.f=full,d.si=6thin}{t}
\end{fmfgraph*}
\end{fmffile}
\caption{$1/N$ corrections to the insertion of $Q^a$ into the $z$
propagator.}\label{diagM}
\end{figure}
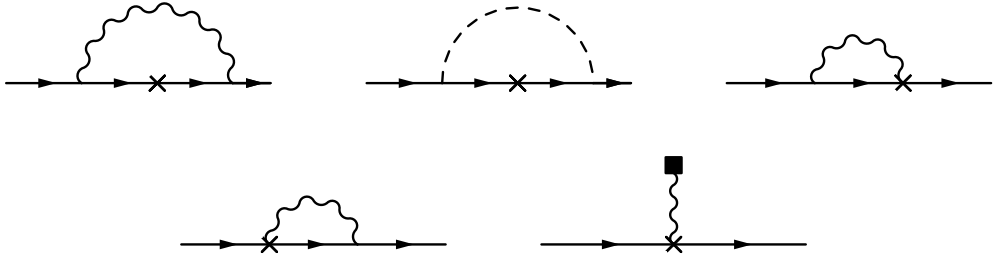

The $1/N$ corrections to the insertion of $Q^a$ into the $z$
propagator are shown in Fig. \ref{diagM}. %\vspace{0.5cm}

Again, only the last one of these couples to the impurity and breaks
translational invariance, so,
\begin{equation}
\label{Mapqpcor} Q^a_{\alpha \beta}(\vec{p},\vec{q},\vec{p}',\omega,
\omega') \stackrel{\vec{q} \neq \vec{p}'-\vec{p}}{=} -2
\frac{1}{L^2} {\cal D}_{\tau\tau}(\vec{q} + \vec{p} - \vec{p}',0)
T^a_{\alpha \beta} + O(1/N^2)
\end{equation}

Combining (\ref{Ocor}),(\ref{Rescor}),(\ref{Mapqpcor}),
\begin{equation}
\label{Mqresult}\langle \alpha |Q^a(\vec{q})|\beta\rangle = -
\left(\delta_{\vec{q},0} + (1- \delta_{\vec{q},0}) \frac{1}{\theta
L} (1 + \frac{4 m^2_0}{\vec{q}^2}) {\cal
D}_{\tau\tau}(\vec{q},0)\right)T^a_{\alpha \beta} + O(1/N^2)
\end{equation}

The calculation of $1/N$ corrections to result (\ref{Naqlead}) for
impurity induced staggered magnetization $n^a(x)$ proceed in the
same fashion. The corrections to insertion of $n^a(x)$ into the $z$
propagator are given by the first two diagrams in Fig. \ref{diagM}
(except now the cross stands for $n^a$ insertion). None of these
break translational invariance (as the last diagram in Fig.
\ref{diagM} is present only for $Q^a$, but not for $n^a$).
Therefore,
\begin{equation}
\label{Nqresult} \langle \alpha| n^a(\vec{q})|\beta\rangle =
\frac{1}{2 m_0} \left(\delta_{\vec{q},0} (1 + O(1/N)) +
(1-\delta_{\vec{q},0}) \frac{1}{L^2} \frac{4 m_0}{\vec{q}^2}{\cal
D}_{\tau\tau}(\vec{q},0)\right) T^a_{\alpha \beta} + O(1/N^2)
\end{equation}
Note again that in the case of $\langle \alpha |n^a(\vec{q})|\beta
\rangle$ we have computed the $1/N$ corrections only to $\vec{q}
\neq 0$. Unlike the case of uniform magnetization, here the $N=
\infty$ result for $\langle \alpha|n^a(\vec{q} = 0)|\beta\rangle$ is
expected to receive corrections.

\begin{figure}[t]
\begin{center}
\includegraphics[width=0.5\textwidth]{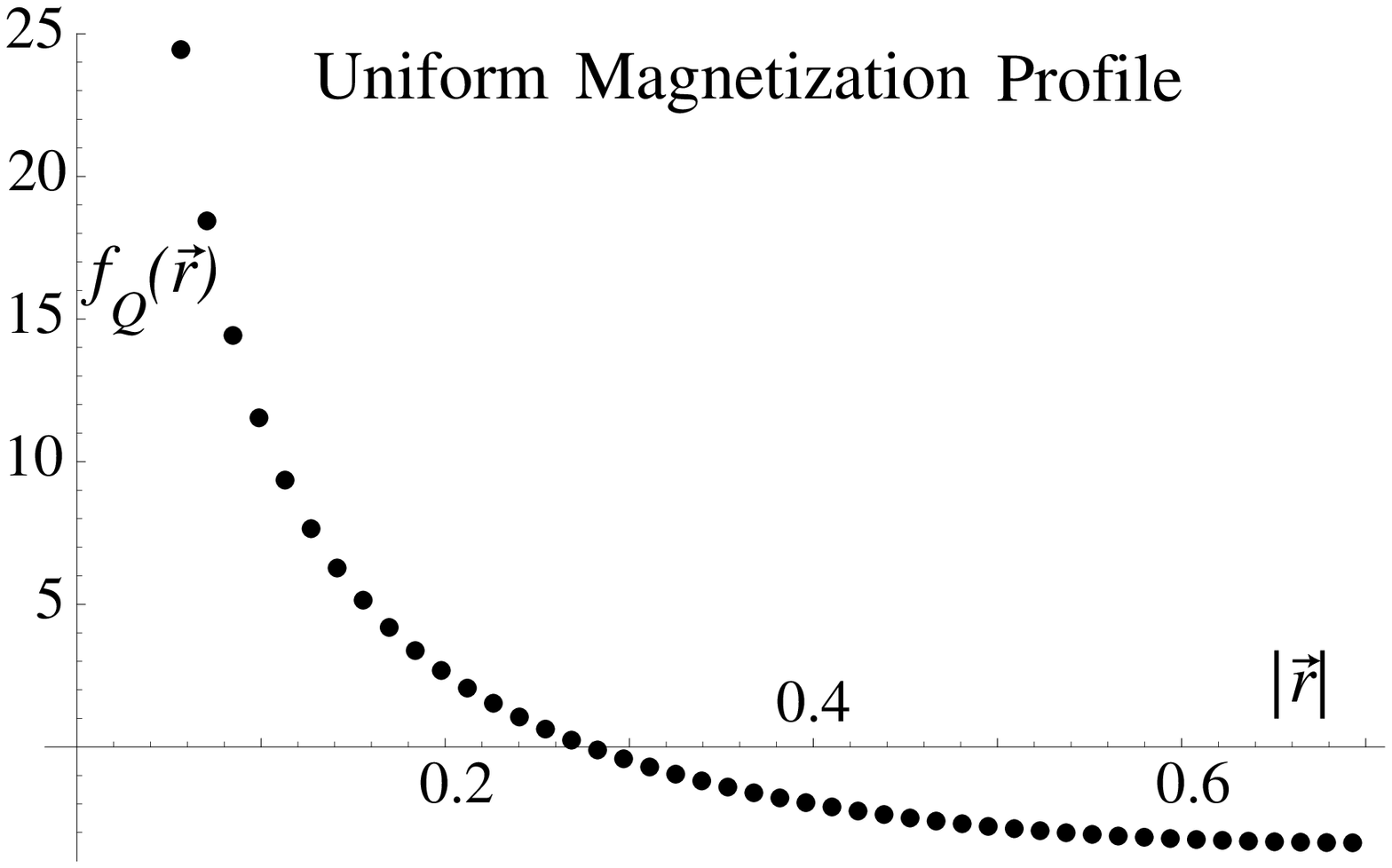}%, width = 0.45\textwidth]{M1ddiag}
\includegraphics[width = 0.5\textwidth]{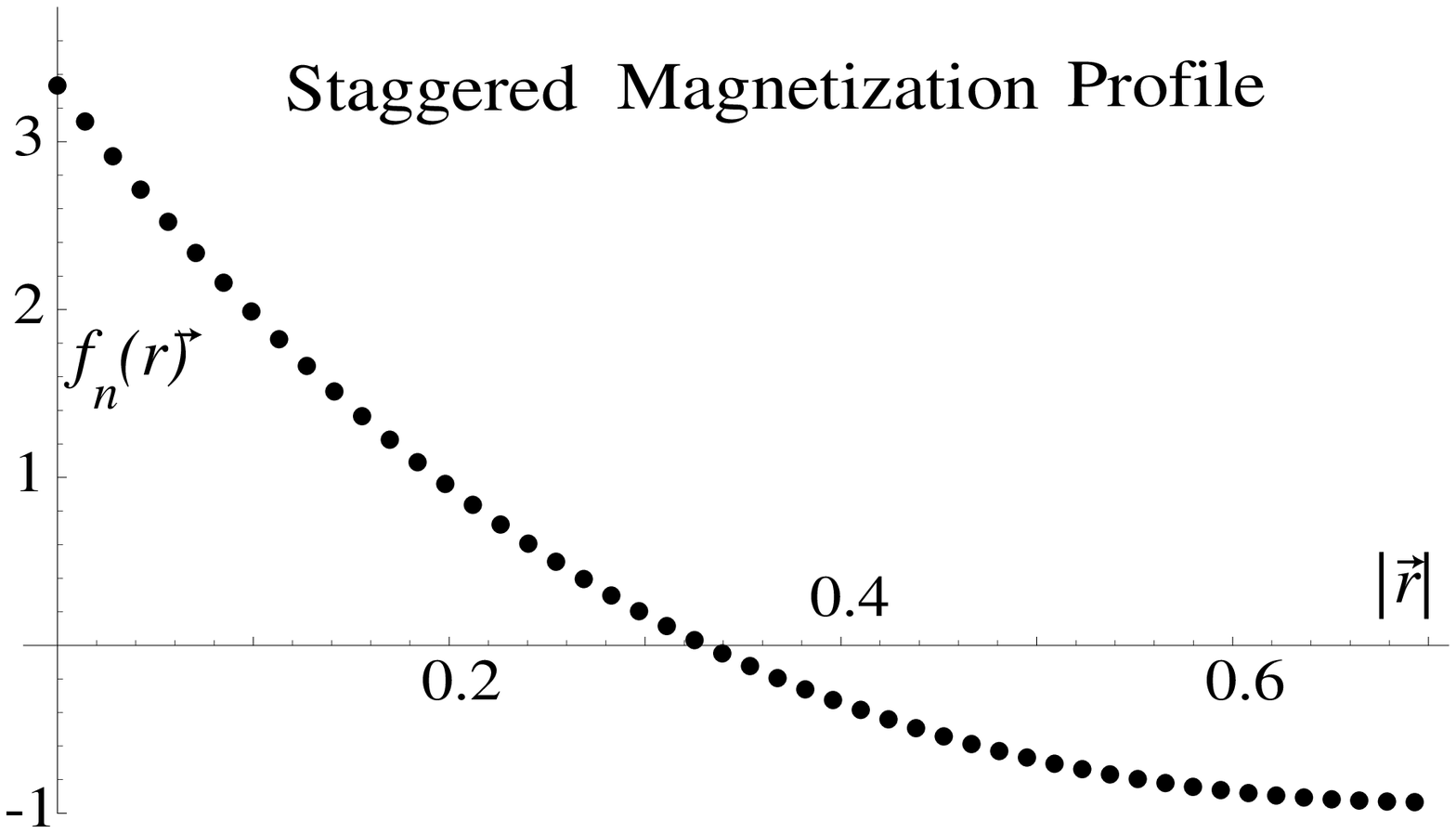}
\caption{Uniform (a) and Staggered (b) magnetization distribution
functions $f_Q(\vec{r})$, $f_n(\vec{r})$ plotted along the torus
diagonal.} \label{diagplots}
\end{center}
\end{figure}

Thus, the scaling functions,
\begin{eqnarray}
\label{PhiMr} \Phi_Q(\vec{x}/L) &=& -1 - \frac{1}{\theta L}
\sum_{\vec{q} \neq 0} (1 + \frac{4
m^2_0}{\vec{q}^2}) {\cal D}_{\tau\tau}(\vec{q},0) e^{i \vec{q} \vec{x}} + O(1/N^2)\\
\label{PhiNr}\Phi_n(\vec{x}/L) &=& \frac{1}{2 \theta}  + c_1 +
\frac{1}{L^3} \sum_{\vec{q} \neq 0} \frac{2}{\vec{q}^2 } {\cal
D}_{\tau\tau}(\vec{q},0)e^{i \vec{q} \vec{x}} + O(1/N^2)
\end{eqnarray}
where $c_1$ is an $\vec{x}$-independent constant of order $1/N$
($c_1$ should be also independent of $\Lambda$; we have not verified
this fact as we did not compute the $1/N$ corrections to
$\langle\alpha|n^a(\vec{q} = 0)|\beta\rangle$). We may write,
\begin{eqnarray} \Phi_Q(\vec{r}) &=& -(1 + \frac{1}{N} f_Q(\vec{r})) + O(1/N^2)\\ \Phi_n(\vec{r})
&=& \frac{1}{2 \theta} (1 + 2 c_1 \theta + \frac{1}{N} f_n(\vec{r}))
+ O(1/N^2)\end{eqnarray} We have evaluated the functions $f_Q$,
$f_n$ numerically and plotted them along the diagonal of our spatial
torus in Fig. \ref{diagplots}.

Now, we would like to find the $\vec{q} \to \infty$, $\vec{x} \to 0$
asymptotes of (\ref{PhiMr}), (\ref{PhiNr}). For this purpose, we may
replace the finite box propagator ${\cal D}_{\tau\tau}(q)$ by the
infinite box propagator,
\begin{equation} {\cal
D}_{\tau\tau}(\vec{q},0) \stackrel{\vec{q} \to \infty}{\to}
\frac{1}{N A} \frac{1}{|\vec{q}|}
\end{equation}
Writing, $\Phi_{Q,n}(\vec{x}/L) = \frac{1}{L^2} \sum_{\vec{q}}
\Phi_{Q,n} (\vec{q}) e^{i \vec{q} \vec{x}}$,
\begin{eqnarray}
\Phi_Q(\vec{q}) &\stackrel{\vec{q} \to \infty}{\to}&
-\frac{1}{N A \theta} \frac{L}{|\vec{q}|} + O(1/N^2)\\
\Phi_n(\vec{q}) &\stackrel{\vec{q} \to \infty}{\to}& \frac{2}{N
A}\frac{1}{L |\vec{q}|^3} + O(1/N^2)
\end{eqnarray}
Fourier
transforming,
\begin{eqnarray}
\Phi_Q(\vec{r}) &\stackrel{|\vec{r}| \to
0}{\to}& -\frac{1}{2 \pi \theta N A} \frac{1}{|\vec{r}|} + O(1/N^2)\\
\Phi_n(\vec{r}) &\stackrel{|\vec{r}| \to 0}{\to}& \frac{1}{2 \theta}
+ c_2 + O(1/N^2)
\end{eqnarray}
where $c_2$ is a constant of order $1/N$. %Thus, \begin{eqnarray}
%\langle \alpha | M^a(\vec{q})|\beta\rangle \stackrel{\vec{q} \to
%\infty}{\to}
%-\frac{1}{N A \theta L} \frac{1}{|\vec{q}|}\, T^a_{\alpha \beta}\\
%\langle \alpha | N^a(\vec{q})|\beta\rangle \stackrel{\vec{q} \to
%\infty}{\to} \frac{2}{L^2 N A} \frac{1}{|\vec{q}|^3}\, T^a_{\alpha
%\beta}\end{eqnarray} Fourier transforming, \begin{eqnarray} \langle
%\alpha|M^a(\vec{x})|\beta \rangle &\stackrel{\vec{x} \to 0}{\to}&
%-\frac{1}{2 \pi \theta N A L} \frac{1}{|\vec{x}|} T^a_{\alpha \beta}
%+ O(1/N^2)\\ \langle \alpha|N^a(\vec{x})|\beta\rangle
%&\stackrel{\vec{x} \to 0}{\to}& (\frac{1}{2 \theta L} + c_1)
%T^a_{\alpha \beta} + O(1/N^2)\end{eqnarray} where $c_1$ is an
%$\vec{x}$-independent quantity of order $1/N$.

Thus, we conclude that,
\begin{equation}
\label{impexp} \Delta^Q_{\rm imp} = -1 + O(1/N) \quad \Delta^n_{\rm
imp} = O(1/N^2)
\end{equation}
which is consistent with the relation between impurity exponents
(\ref{impdimrel}). Note that the present calculation shows that
$\Delta^n_{\rm imp}$ is zero to order $1/N$. We shall verify this
fact in a different way in Section~\ref{sec:neel}, and compute
$\Delta^n_{\rm imp}$ to order $1/N^2$.

Moreover, the ratio,
\begin{equation}
\label{ratio}\frac{c_Q}{c_n} = -\frac{1}{\pi N A} + O(1/N^2)
\end{equation}
is independent of regularization at this order in $N$.

\subsection{N\'eel Phase}
\label{sec:neel}

In this section, we compute the uniform and staggered magnetization
in the presence of an impurity of charge $Q$ in the symmetry broken
phase, $g < g_c$. We work in infinite volume. We develop the $1/N$
expansion around the symmetry broken vacuum,
\begin{equation}
\langle z_1 \rangle = \frac{1}{\sqrt{2}} v
\end{equation} Note that
in general $v$ is not a gauge invariant quantity. However, this fact
does not manifest itself at the order at which we are working. To
leading order in $N$,
\begin{equation}
\frac{1}{2} v^2 = \frac{1}{g}
- \frac{1}{g_c}
\end{equation}
Note that $v^2 \sim O(N)$. Moreover,
we take $Q \sim O(1)$ in $N$.

We now must quantize our theory around the symmetry broken state. We
write,
\begin{equation}
z_1 = \frac{1}{\sqrt{2}} (h + v + i \phi),\quad z_\alpha =
\pi_\alpha,\, \alpha = 2\,..\, N\end{equation} We work in the
so-called $R_\xi$ gauge, in which the mixing between the goldstone
$\phi$ and the photon $A_{\mu}$ is absent, at the expense of
introducing a ghost field $c$. In what follows, we have eliminated
the mixing only to leading order in $1/N$.  This is achieved by
using the gauge-fixing condition,
\begin{equation}
\d_{\mu} A_{\mu} = \xi v K^{-1} \phi + w
\end{equation} where the
action for the auxillary field $w$, which appears in the
Fadeev-Poppov formalism, is
\begin{equation}
{\cal S}_w = \frac{1}{2 \xi} \int dx dy w(x) K(x-y) w(y)
\end{equation} Here, $K(x - y)$ is the photon polarization function given by Eq.
(\ref{Kc}). Similarly, in what follows $\Pi(x-y)$ is the $\lambda$
self energy given by Eq. (\ref{Pic}).

At the end of the day, the action one obtains is,
\begin{eqnarray}
{\cal S}_{\xi} &=& \frac{1}{2} \int dx dy \,A_{\mu} (x) \left(K_{\mu
\nu} (x-y) -\frac{1}{\xi}
\d_{\mu} \d_{\nu} K(x-y) + \delta_{\mu \nu} v^2\right) A_{\nu}\nn \\
&+& \frac{1}{2} \int dx dy \, \phi(x) \left( -\d^2 \delta(x-y) + \xi
v^2 K^{-1} (x-y) \right)\phi(y) +\frac{1}{2} \int dx dy
\lambda(x)\Pi(x-y) \lambda(y)\nn\\ &+& \int dx dy \Big( \bar{c}
\left(-\d^2 \delta(x-y) + \xi v^2 K^{-1}(x-y) \right)c(y) + \xi v
\bar{c}(x) K^{-1}(x-y) h(y) c(y)\Big)\nn\\&+& \int dx \left(|D_{\mu}
\pi|^2 +  \frac{1}{2} (\d_{\mu} h)^2 + i v \lambda h + (\phi
\d_{\mu} h - \d_{\mu} \phi h) A_{\mu} + (v h + \frac{1}{2} h^2 +
\frac{1}{2} \phi^2) A^2_{\mu}\right)\nn\\ &+& \int dx \left( i
\lambda |\pi|^2 + \frac{1}{2} i \lambda (\phi^2 + h^2) \right)
\end{eqnarray}
As usual, we avoid double counting by dropping any diagrams, which
are already included in the dynamically generated $N = \infty$
self-energies for $A_{\mu}$, $\lambda$ etc. The propogators for our
fields are shown in Fig. \ref{fprops}. Note that in the N\'eel
phase, we get mixing between the $\lambda$ and $h$ fields.

\begin{figure}[h]
\begin{flushleft}
\begin{fmffile}{PropsCond}
\parbox{50mm}{
\begin{fmfgraph*}(100,40)
\fmfset{arrow_len}{2.5mm} \fmfleft{l} \fmfright{r} \fmf{wiggly}{l,r}
\fmflabel{$A_\mu$}{l} \fmflabel{$A_\nu$}{r}
\end{fmfgraph*}} $\quad {\cal D}_{\mu \nu}(p) = \frac{1}{p^2 K(p) +
v^2}\left(\delta_{\mu \nu} + \frac{(\xi -1) K(p) p_{\mu}
p_{\nu}}{p^2 K(p) + \xi v^2}\right)$\\
\parbox{50mm}{\begin{fmfgraph*}(100,40) \fmfset{arrow_len}{2.5mm} \fmfleft{l}
\fmfright{r} \fmf{zigzag}{l,r} \fmflabel{$\phi$}{l}
\fmflabel{$\phi$}{r}
\end{fmfgraph*}} $\quad D_{\phi}(p) = \frac{1}{p^2+ \xi v^2
K^{-1}(p)}$\\
\parbox{50mm}{\begin{fmfgraph*}(100,40) \fmfset{arrow_len}{2.5mm} \fmfleft{l}
\fmfright{r} \fmf{dots}{l,r} \fmflabel{$\bar{c}$}{l}
\fmflabel{$c$}{r}
\end{fmfgraph*}} $\quad D_{c}(p) = \frac{1}{p^2+ \xi v^2
K^{-1}(p)}$\\
\parbox{50mm}{\begin{fmfgraph*}(100,40) \fmfset{arrow_len}{2.5mm} \fmfleft{l}
\fmfright{r} \fmf{plain}{l,r} \fmflabel{$\pi^*_{\alpha}$}{l}
\fmflabel{$\pi_{\beta}$}{r}
\end{fmfgraph*}} $\quad {D_{\pi}}_{\alpha \beta}(p) = \frac{\delta_{\alpha
\beta}}{p^2}$\\
\parbox{50mm}{\begin{fmfgraph*}(100,40) \fmfset{arrow_len}{2.5mm} \fmfleft{l}
\fmfright{r} \fmf{plain,width=1thick}{l,r} \fmflabel{$h$}{l}
\fmflabel{$h$}{r}
\end{fmfgraph*}} $\quad D_{h}(p) = \frac{\Pi(p)}{p^2 \Pi(p) +
v^2}$\\
\parbox{50mm}{\begin{fmfgraph*}(100,40) \fmfset{arrow_len}{2.5mm} \fmfleft{l}
\fmfright{r} \fmf{dashes}{l,r} \fmflabel{$\lambda$}{l}
\fmflabel{$\lambda$}{r}
\end{fmfgraph*}} $\quad D_{\lambda}(p) = \frac{p^2}{p^2 \Pi(p) +
v^2}$\\
\parbox{50mm}{\begin{fmfgraph*}(100,40) \fmfset{arrow_len}{2.5mm} \fmfleft{l}
\fmfright{r} \fmf{plain,width=1thick}{l,v1}\fmf{dashes}{v1,r}
\fmflabel{$h$}{l} \fmflabel{$\lambda$}{r}
\end{fmfgraph*}} $\quad D_{h \lambda}(p) = \frac{-i v}{p^2 \Pi(p) +
v^2}$
\end{fmffile}
\end{flushleft}
\caption{Propagators in the N\'eel phase.}\label{fprops}
\end{figure}

Now, having set up the perturbation theory, we wish to compute,
$\langle Q^a(\vec{x})\rangle$,  $\langle n^a(\vec{x}) \rangle$.
Utilizing the pattern of spontaneous symmetry breaking, $U(N) \to
U(N-1)$ (here we look only at global symmetry), one can show that,
\begin{equation}
\langle n^a \rangle = T^a_{11} \langle n^0 \rangle
\end{equation}
where $n^0 = z^{\dagger} T^0 z$ and $T^0$ is any generator of
SU($N$) with $T^0_{11} = 1$. Similarly for $Q^a$. For definiteness,
we may choose $T^0_{11} = 1$, $T^0_{1 \alpha} = T^0_{\alpha 1} = 0$,
$T^0_{\alpha \beta} = -\frac{1}{N-1} \delta_{\alpha \beta}$,
$\alpha, \beta = 2\,..\,N$.

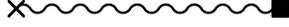
\begin{figure}[t]
\begin{fmffile}{A0tad}
\fmfset{arrow_len}{2.5mm}
\begin{fmfgraph*}(100,40)
\fmfleft{l} \fmfright{r} \fmf{wiggly}{l,r}
\fmfv{d.sh=cross,d.f=empty,d.si=8thin}{l}
\fmfv{d.sh=square,d.f=full,d.si=6thin}{r}
\end{fmfgraph*}
\end{fmffile}
\caption{Leading contribution to uniform magnetization in the
symmetry broken phase.}\label{A0tad}
\end{figure}

Let's start with computing the uniform magnetization.
\begin{equation}
Q^0
 = \frac{N}{N-1} j^1_{\tau} - \frac{1}{N-1} j_{\tau}
\end{equation}
where
\begin{equation}
j^1_{\tau} = z^{\dagger}_1 D_{\tau} z_1 - (D_{\tau} z_1)^{\dagger}
z_1
\end{equation}
and $j_{\tau}$ is the $U(1)$ charge density discussed in the
appendix, see Eq. (\ref{jU1}). By equation of motion (\ref{screen}),
\begin{equation}
\langle j_{\tau}(\vec{x}) \rangle = - J^{{\rm
ext}}_{\tau}(\vec{x}) = -Q \delta^2(\vec{x})
\end{equation}
So, it remains to compute $\langle j^1_{\tau}(\vec{x})\rangle$.
Expanding $j^1_\tau$ in terms of $\phi$, $h$ and $A_{\mu}$,
\begin{equation}
\label{j1exp} j^1_{\tau} = -i v^2 A_{\tau} + i v (\d_{\tau} \phi - 2
A_{\tau} h) + i (h \d_{\tau} \phi - \phi \d_{\tau} h - A_{\tau} (h^2
+ \phi^2))
\end{equation}
In the $1/N$ expansion the leading
contribution to $\langle j^1_{\tau} \rangle$ is of $O(1)$ and comes
from the first term on the r.h.s of (\ref{j1exp}), see Fig.
\ref{A0tad}.
\begin{equation}
\langle j^1_{\tau}(\vec{p})\rangle = - Q v^2 {\cal
D}_{\tau\tau}(\vec{p}, 0) = - Q v^2 \frac{1}{\vec{p}^2 K(\vec{p}) +
v^2}
\end{equation}
Thus, $\langle Q^0(\vec{p})\rangle = \langle j^1_{\tau}(\vec{p})
\rangle + O(1/N)$. Fourier transforming,
\begin{equation}
\langle Q^0(\vec{x})\rangle = - \frac{Q v^2}{2 \pi N A}
\frac{1}{|\vec{x}|} + \frac{Q v^4}{4 N^2 A^2} \left(
\bm{H}_0\left(\frac{v^2|\vec{x}|}{N A}\right) -
Y_0\left(\frac{v^2|\vec{x}|}{N A}\right)\right) + O(1/N)\,,
\end{equation}
where $\bm{H}_0$ is the Struve function and $Y_0$ is the Bessel
function. Taking the short and long distance asymptotes,
\begin{equation}
\langle Q^0(\vec{x})\rangle \stackrel{\vec{x} \to 0} {\to} -\frac{Q
v^2}{2 \pi N A} \frac{1}{|\vec{x}|}
\end{equation}
\begin{equation}
\langle Q^0(\vec{x})\rangle \stackrel{\vec{x} \to \infty}{\to} -
\frac{Q N A}{2 \pi v^2} \frac{1}{|\vec{x}|^3}
\end{equation}
The long distance decay is a consequence of the Goldstone physics of
the spin waves, and the $1/|\vec{x}|^3$ decay is expected to be
exact. At short distances, we have the physics of the critical
point, and the exponent will have corrections at higher order. From
the present result we can conclude that the impurity exponent
\begin{equation}
\Delta^Q_{\rm imp} = -1 + O(1/N),
\end{equation}
which is
consistent with the result obtained at the critical point
(\ref{impexp}).

\begin{figure}[t]
\begin{fmffile}{stagcond}
\fmfset{arrow_len}{2.5mm}
\begin{fmfgraph*}(100,60)
\fmfleft{l} \fmfright{ru,rd}
\fmf{plain,width=1thick,label=$a)$,label.dist=20}{l,v1}
\fmf{wiggly}{v1,ru} \fmf{wiggly}{v1,rd}
\fmfv{d.sh=cross,d.f=empty,d.si=8thin}{l}
\fmfv{d.sh=square,d.f=full,d.si=6thin}{ru}
\fmfv{d.sh=square,d.f=full,d.si=6thin}{rd}
\end{fmfgraph*}\\
\vspace{0.75cm}
\begin{fmfgraph*}(100,80)
\fmfleft{l} \fmfright{ru,rd}
\fmf{plain,width=1thick,label=$b)$,label.dist=20,label.side=right}{l,v1}
\fmf{dashes}{v1,v2}\fmf{plain_arrow,right,tension=0.5}{v3,v2,v3}
\fmf{wiggly}{v3,ru}\fmf{wiggly}{v3,rd}
\fmfv{d.sh=cross,d.f=empty,d.si=8thin}{l}
\fmfv{d.sh=square,d.f=full,d.si=6thin}{ru}
\fmfv{d.sh=square,d.f=full,d.si=6thin}{rd}
\end{fmfgraph*}
\hspace{1cm}
\begin{fmfgraph*}(100,80)
\fmfleft{l} \fmfright{ru,rd}
\fmf{plain,width=1thick,label=$c)$,label.dist=20}{l,v1}
\fmf{dashes}{v1,v2}\fmf{wiggly}{v3,ru}\fmf{wiggly}{v4,rd}
\fmf{plain_arrow,right}{v3,v4} \fmf{plain_arrow,right}{v4,v2}
\fmf{plain_arrow,right}{v2,v3}
\fmfv{d.sh=cross,d.f=empty,d.si=8thin}{l}
\fmfv{d.sh=square,d.f=full,d.si=6thin}{ru}
\fmfv{d.sh=square,d.f=full,d.si=6thin}{rd}
\end{fmfgraph*}
\end{fmffile}
\caption{Leading $\vec{x}$-dependent contribution to staggered
magnetization in the symmetry broken phase.}\label{stagcond}
\end{figure}
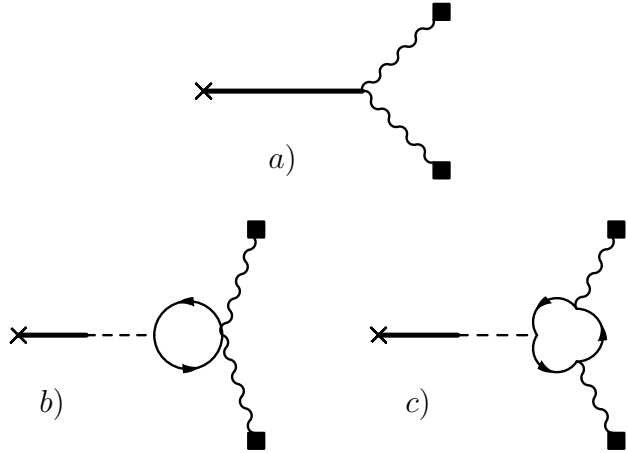

Now, let's discuss the staggered magnetization,
\begin{equation}
n^0 = \frac{N}{N-1} z^{\dagger}_1 z_1 - \frac{1}{N-1} z^{\dagger}
z\end{equation} By equations of motion,
\begin{equation} z^{\dagger}
z = \frac{1}{g}
\end{equation}
thus,
\begin{equation}
n^0 = \frac{N}{N-1} z^{\dagger}_1 z_1 -\frac{1}{(N-1) g}
\end{equation}
 and
\begin{equation}
z^{\dagger}_1 z_1 = \frac{1}{2} v^2 + v h + \frac{1}{2}(h^2 +
\phi^2)
\end{equation}
Thus, at leading order, $\langle z^{\dagger}_1 z_1 (\vec{x}) \rangle
= \frac{1}{2} v^2$, and
\begin{equation}
\langle n^0(\vec{x}) \rangle = \frac{1}{2} v^2 + O(1)
\end{equation}
Moreover, the $\vec{x}$-dependent corrections to $\langle
n^0(\vec{x}) \rangle$ come only at $O(1/N)$, with diagrams of Fig.
\ref{stagcond} (the part of $n^0$ which contributes at this order,
denoted by $\times$, is $v h$).

We will discuss the diagrams in Fig. \ref{stagcond} shortly. For
now, we can conclude that,
\begin{equation}
\Delta^n_{\rm imp} = O(1/N^2)
\end{equation}
in agreement with the result (\ref{impexp})
obtained at the critical point. Moreover, we can now compute the
ratio,
\begin{equation}
\frac{c_Q}{c_n} = -\frac{Q}{\pi N A} + O(1/N^2)
\end{equation}
which exactly agrees with the result obtained at the critical point
(\ref{ratio}) for $Q = 1$. Notice, that this is a highly nontrivial
check of the OPE (\ref{OPE}) as $\langle Q^a \rangle$, $\langle n^a
\rangle$ depend on $v$ in the N\'eel phase and on $L$ at the
critical point. Nevertheless, all the dependence on the IR scale
cancels out in the ratio $c_Q/c_n$, which is constant throughout the
scaling regime.

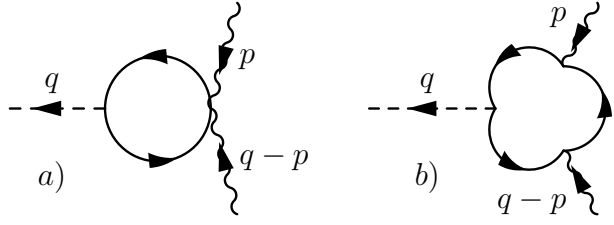
\begin{figure}[t]\begin{fmffile}{Gmn}
\begin{fmfgraph*}(100,80)
\fmfleft{l} \fmfright{ru,rd} \fmf{dashes_arrow,label=$q$}{v2,l}
\fmf{plain_arrow,right,tension=0.5}{v3,v2,v3}
\fmf{phantom_arrow,label=$q-p$,label.side=right}{ru,v3}\fmf{wiggly}{ru,v3}\fmf{wiggly}{rd,v3}\fmf{phantom_arrow,label=$p$,label.side=left}{rd,v3}
\fmffreeze
\fmf{phantom,label=$a)$,label.side=right,label.dist=20}{l,v2}
\end{fmfgraph*}
\hspace{1cm}
\begin{fmfgraph*}(100,80)
\fmfleft{l} \fmfright{ru,rd}
\fmf{dashes_arrow,label=$q$}{v2,l}\fmf{wiggly}{v3,ru}\fmf{phantom_arrow,label=$q-p$}{ru,v3}\fmf{wiggly}{v4,rd}\fmf{phantom_arrow,label=$p$}{rd,v4}
\fmf{plain_arrow,right}{v3,v4} \fmf{plain_arrow,right}{v4,v2}
\fmf{plain_arrow,right}{v2,v3} \fmffreeze
\fmf{phantom,label=$b)$,label.side=right,label.dist=20}{l,v2}
\end{fmfgraph*}
\end{fmffile}
\caption{Leading contribution to the three point vertex of
$A_{\mu}$, $A_{\nu}$ and $\lambda$ fields, $\Gamma^{\mu \nu}(q, p,
q-p)$.}\label{Gmn}
\end{figure}

Coming back to the diagrams in Fig. \ref{stagcond},
\begin{equation}
\langle n^0(\vec{q})\rangle \stackrel{\vec{q} \neq 0}{=} Q^2 v^2\!
D_h(\vec{q},0)\!\! \int\!\!\frac{d^{D-1} p}{(2
\pi)^{D-1}}\!\left(\!1\! +\! \frac{i}{2}\Pi^{-1}(\vec{q},0)
\Gamma^{\tau\tau}(\vec{q},0,\vec{p},0,\vec{q}-\vec{p},0)\!\right)\!
{\cal D}_{\tau\tau}(\vec{p},0)\! {\cal
D}_{\tau\tau}(\vec{q}-\vec{p},0)
\end{equation} We keep the
space-time dimension $D$ arbitrary in what follows, as we wish to
compare our result for $\Delta^n_{\rm imp}$ obtained in the $1/N$
expansion, with the result obtained using $\epsilon$
expansion\cite{ImpurityKSBC}. Here, $\Gamma^{\mu \nu}(q, p,q-p)$ is
the lowest order contribution to the $A_{\mu}$, $A_{\nu}$, $\lambda$
vertex, given by the sum of the loops in Fig. \ref{Gmn}.
The diagram in Fig. \ref{Gmn} a) is given by,
\begin{equation}
\Gamma^{\mu \nu}_1(q,p,q-p) = 2 i \delta_{\mu \nu}
\Pi(q)\end{equation}
Thus, diagrams in Fig. 7 a) and b) cancel (by
the way, these diagrams are individually UV divergent for $D \leq
3$). So, calling the diagram in Fig. \ref{Gmn} b), $\Gamma^{\mu
\nu}_2(q,p,q-p)$,

\begin{equation}
\label{N0short} \langle n^0(\vec{q})\rangle \stackrel{\vec{q} \neq
0}{=} Q^2 v^2 D_h(\vec{q},0) \int\frac{d^{D-1} p}{(2
\pi)^{D-1}}\frac{i}{2}\Pi^{-1}(\vec{q},0)
\Gamma^{\tau\tau}_2(\vec{q},0,\vec{p},0,\vec{q}-\vec{p},0) {\cal
D}_{\tau\tau}(\vec{p},0) {\cal D}_{\tau\tau}(\vec{q}-\vec{p},0)
\end{equation}
Evaluating $\Gamma^{\mu \nu}_2$,
\begin{eqnarray}
\Gamma^{\mu \nu}_2(q,p,q-p) &=& -2 i N \int \frac{d^D l}{(2 \pi)^D}
\frac{(2l-p)_{\mu} (2l-p-q)_{\nu}}{l^2 (l-p)^2 (l-q)^2} \nn\\&=&
-\frac{4 i N \Gamma(2-D/2)}{(4 \pi)^{D/2}} \int dx_1 dx_2 dx_3
\delta(1-x_1-x_2-x_3) (\Delta^2)^{D/2-2}\nn\\&&\Big(\delta_{\mu \nu}
 +\, \frac{(4-D)(2 x_1 q + (2 x_2-1) p)_{\mu} ((2 x_1-1)q + (2 x_2
-1)p)_{\nu}}{4 \Delta^2}\Big)\nn\\
\end{eqnarray}
where,
\begin{equation}
\Delta^2 = x_1(1-x_1) q^2 + x_2 (1-x_2) p^2 - 2 x_1
x_2\, p\cdot q
\end{equation}
We are interested only in
$\Gamma^{\tau\tau}_2$, with $p^0 = q^0 = 0$. Thus,
\begin{eqnarray}
\label{G00}
\Gamma^{\tau\tau}_{2}(\vec{q},0,\vec{p},0,\vec{q}-\vec{p},0)
=-\frac{4 i N \Gamma(2-D/2)}{(4 \pi)^{D/2}}\int dx_1 dx_2 dx_3
\delta(1-x_1-x_2-x_3) (\Delta^2)^{D/2-2}\nn\\
\end{eqnarray} For $|\vec{p}| \gg
|\vec{q}|$,
$\Gamma^{\tau\tau}_{2}(\vec{q},0,\vec{p},0,\vec{q}-\vec{p},0) \sim
|\vec{p}|^{D-4}$, so for $\vec{p} \to \infty$ the integrand in Eq.
(\ref{N0short}) behaves as $|\vec{p}|^{-D}$ and the integral is UV
convergent.

We now attempt to understand the behaviour of (\ref{N0short}) for
$\vec{q} \to \infty$, from which we should be able to extract the
impurity anomalous dimension $\Delta^n_{\rm imp}$. For this purpose,
we may set $v = 0$ in the propagators $D_h(\vec{q},0)$, ${\cal
D}_{\tau\tau}(\vec{p},0)$, ${\cal D}_{\tau\tau}(\vec{q}-\vec{p},0)$
(this does not introduce any IR divergences).
\begin{equation}
\langle n^0(\vec{q})\rangle \stackrel{\vec{q} \to \infty}{=}
\frac{Q^2 v^2}{N^3 A^2 B} |\vec{q}|^{2-D}\int\frac{d^{D-1} p}{(2
\pi)^{D-1}} \frac{i}{2}
\Gamma_2^{\tau\tau}(\vec{q},0,\vec{p},0,\vec{q}-\vec{p},0)\frac{1}{|\vec{p}|^{D-2}}\frac{1}{|\vec{p}-\vec{q}|^{D-2}}
\end{equation}

Let us first discuss the limit $D = 4-\epsilon$, $\epsilon \to 0$.
In this regime, to leading order in $\epsilon$,
\begin{equation}
\Gamma^{\tau\tau}_2(\vec{q},0,\vec{p},0,\vec{q}-\vec{p},0) = -2 i N
\frac{1}{(4 \pi)^{2}}\Gamma(2-D/2)
\end{equation}
and,
\begin{equation}
\langle n^0(\vec{q})\rangle \stackrel{\vec{q} \to \infty}{=}
\frac{72 \pi^4 \epsilon^2 Q^2 v^2}{N^2} \frac{1}{|\vec{q}|^3}
\end{equation}
Fourier transforming,
\begin{equation} \langle n^0(\vec{x})\rangle \stackrel{\vec{x} \to
0}{=} \frac{1}{2} v^2 + c_3 - \frac{36 \pi^2 \epsilon^2 Q^2}{N^2}
v^2 \log(v |\vec{x}|) + c_4 + O(1/N^2)
\end{equation} where $c_3$,
$c_4$ do not depend on $\vec{x}$ and are of order $1$ and $1/N$
respectively. Thus, to leading order in $1/N$, $\epsilon$,
\begin{equation}
\Delta^n_{\rm imp} = -\frac{72 \pi^2 Q^2 \epsilon^2}{N^2}
\end{equation}
in agreement with the calculations of Ref.
\onlinecite{ImpurityKSBC}, where the impurity exponents were
obtained by performing the impurity operator renormalizations as
summarized in Section~\ref{sec:scaling}.\footnote{Note that in the
$\epsilon$ expansion of Ref.~\onlinecite{ImpurityKSBC} only the
analogue of the diagram in Fig. \ref{stagcond} a) appears, while the
diagrams in Figs. \ref{stagcond} b), c) do not appear at leading
order in $\epsilon$, as they are higher order in coupling constant.
Nevertheless in the $1/N$ expansion, we saw that the answer comes
entirely from the diagram in Fig. \ref{stagcond} c), with diagrams
in Fig. \ref{stagcond} a) and Fig. \ref{stagcond} b) canceling for
all $D$. The reason is the following: in the $1/N$ expansion all
diagrams in Fig. \ref{stagcond} are individually of same order in
$\epsilon$. Moreover, to leading order in $\epsilon$, the diagrams
b) and c) cancel, so a) = - b) $\stackrel{\epsilon \to 0}{=}$ c). In
the $\epsilon$ expansion, this fact is foreseen in advance: the
$1/\epsilon$ pole must cancel between diagrams b) and c) (the
$4$-point diagram with two photons and two scalars is not
divergent). Thus, we can obtain the answer to leading order in
$\epsilon$ either from a) alone or from c) alone.}

For arbitrary $D$, $\Delta^n_{\rm imp}$ is difficult to calculate
analytically, as $\Gamma^{\tau\tau}$ is no longer a constant.
However, combining Eqs. (\ref{N0short}), (\ref{G00}) and introducing
a new set of Feynman parameters,
\begin{equation}
\langle n^0(\vec{q})\rangle \stackrel{\vec{q} \to \infty}{=}
\frac{Q^2 v^2}{N^2}\frac{1}{|\vec{q}|^{D-1}} f(D)
\end{equation}
where the
numerical constant $f(D)$ is given by,
\begin{eqnarray}
&&f(D) = \frac{1}{A^2 B (4 \pi)^{D-1} \Gamma(D/2-1)^2} \int_0^1 dx_1
\int_0^{1-x_1} dx_2 \int_0^1 dy_1 \int_0^{1-y_1} dy_2
\nn\\&&x_2^{(D-3)/2} (1-x_2)^{D/2-1} y_1^{1-D/2} y_2^{D/2-2}
(1-y_1-y_2)^{D/2-2}\nn\\&& (x_2 (1-x_2)^2 y_2 (1-y_2) + x_1 y_1
((1-x_1)(1-x_2)-2 y_2 x_2 (1-x_2) - y_1 x_1 x_2))^{-\frac12}\nn\\
\end{eqnarray}
Consequently,
\begin{equation}
\langle n^0(\vec{x})\rangle \stackrel{\vec{x} \to 0}{=} \frac{1}{2}
v^2 + c_3 - \frac{2}{(4\pi)^{(D-1)/2} \Gamma((D-1)/2)} f(D)
\frac{Q^2}{ N^2} v^2 \log(v^{2/(D-2)} |\vec{x}|) + c_4 + O(1/N^2)
\end{equation}
and
\begin{equation}
\Delta^n_{\rm imp} =- \frac{4}{(4\pi)^{(D-1)/2} \Gamma((D-1)/2)}
f(D) \frac{Q^2}{ N^2} + O(1/N^3) \label{deltanimp}
\end{equation}

Evaluating $f(D)$ numerically for $D = 3$, \begin{equation}
\Delta^n_{\rm imp} \approx -25.9 \, \frac{Q^2}{N^2} + O(1/N^3)
\label{Deltanumer}\end{equation} We note that we have separately
verified the result (\ref{deltanimp}) by performing the impurity OPE
program as summarized in Section~\ref{sec:scaling}.

\section{Conclusions}
\label{sec:conc}

A recent numerical study \cite{sandvik3} examined the spin
distribution in the vicinity of a non-magnetic impurity in a
double-layer, $S=1/2$ square lattice antiferromagnet at its quantum
critical point. The ground state of the system has total spin
$S=1/2$, and the spin distribution of this $S=1/2$ was found to be
extended across the entire system. Universal scaling forms
(Eqs.~(\ref{Qscale}) and (\ref{nscale})) for the uniform and
staggered spin distributions were postulated\cite{sandvik3}, and
found to be in excellent agreement with the numerical results.

This paper has presented the field-theoretic foundation of the above
results. Using the soft-spin O(3) LGW field theory in
Eq.~(\ref{zlgw}), we found that the universal scaling forms in
Eqs.~(\ref{Qscale}) and (\ref{nscale}) were indeed obeyed in an
expansion in $(3-d)$ (where $d$ is the spatial dimensionality), and
explicit results for the universal scaling functions appear in
Eq.~(\ref{d3}).

Next, we examined a similar non-magnetic impurity in $S=1/2$
antiferromagnets which have a single $S=1/2$ spin per unit cell.
Such antiferromagnets can display a deconfined quantum phase
transition\cite{senthil1,senthil2} between N\'eel and valence bond
solid (VBS) states. An explicit example of such a transition was
found recently in Ref.~\onlinecite{sandvik5}. We expect that such
studies will be extended to include non-magnetic impurities in the
future, and so have provided our theoretical predictions here. The
field theory for this situation in $\mathcal{S}_b^z +
\mathcal{S}_{\rm imp}^z$ in Eqs.~(\ref{sbz},\ref{simpz}). It
describes the dynamics of a SU($N$) spinor field,  $z_\alpha$ (the
spinon), and we obtained its critical properties in a $1/N$
expansion. Projecting onto the total spin $S=1/2$ sector of this
theory (which contains the ground state in the presence of the
impurity) was not straightforward here, and we achieved this by the
relation Eq.~(\ref{matel}). Our results obey scaling forms which
appear in Section~\ref{sec:scaling}. The scaling functions are in
Eqs.~(\ref{PhiMr}) and (\ref{PhiNr}), and are plotted in
Fig.~\ref{diagplots}. The boundary spin exponent for the deconfined
critical point appears in Eqs.~(\ref{deltanimp}),
(\ref{Deltanumer}). We also obtained substantial evidence for the
structure of the operator product expansion near the impurity, and
the fact that the staggered and uniform magnetizations flow to the
same impurity spin operator.

Our study of the deconfined case has so far only examined the
uniform and staggered spin magnetizations near the non-magnetic
impurity. We have not provided here a description of the structure
of the VBS order near the impurity. This will be presented in a
forthcoming paper.

\acknowledgments

We are grateful to A.~Sandvik for valuable discussions. This
research was supported by the NSF grant DMR-0537077.

\appendix

\section{$U(1)$ Charge Density}

Throughout the paper we have concentrated on computing matrix
elements of uniform and staggered magnetization $Q^a(x)$, $n^a(x)$.
However, for the deconfined critical point, it is also interesting
to compute the charge density associated with the $U(1)$ local
symmetry of the $\mathbb{CP}^{N-1}$ model. This charge density is
the zeroth component of the current,
\begin{equation}
\label{jU1} j_{\mu}(x) = z^{\dagger}D_{\mu} z - (D_{\mu}
z)^{\dagger} z
\end{equation}
As we shall see this computation serves as an additional test of our
procedure for projecting onto the single spinon state.

Consider the $\mathbb{CP}^{N-1}$ model coupled to an external
current,
\begin{equation}
\mathcal{S} = \mathcal{S}_b^z + i \int d^3 x A_{\mu} J^{{\rm
ext}}_{\mu}
\end{equation}
As in the rest of the paper, we set $e^2 = \infty$, so that the
gauge field has no bare kinetic term. Then, by equations of motion,
\begin{equation} 0 =
\frac{\delta \mathcal{S}}{\delta A_{\mu}} = i (j^{\mu} + J_{\mu})
\end{equation}
\begin{equation}
\label{screen}j_{\mu} = - J^{{\rm ext}}_{\mu}
\end{equation}
Thus, the dynamical current completely screens (locally!) the
external current. Eq. (\ref{screen}) is an operator identity, and
should, in particular, hold in the ground state of the system with a
single impurity. Let's check this statement in the $1/N$ expansion.

We start from Eq. (\ref{matel}), with $O(x) = j_0(x)$. We write,
 \begin{equation} \langle \alpha |j_0(\vec{x})| \beta \rangle =
\rho(\vec{x}) \delta_{\alpha \beta} \end{equation} with
$\rho(\vec{x}) = \frac{1}{L^2} \sum_{\vec{q}} \rho(\vec{q}) e^{i
\vec{q} \vec{x}}$. The Wilson line term in Eq. (\ref{matel}) can be
incorporated into the action as coupling to an external current,
$J^{{\rm ext}}_{\mu}(\vec{x},\tau) = \delta_{\mu 0}
\delta^3(\vec{x}) \theta({\cal T}/2-\tau) \theta(\tau + {\cal
T}/2)$. At leading order in $1/N$ the numerator of Eq. (\ref{matel})
is given by diagrams shown in Fig. \ref{diagU1}, while the
denominator is given by the bare propagator $D(\vec{x} = 0, {\cal
T})$.

\begin{figure}[h]
\begin{fmffile}{U1}
\fmfset{arrow_len}{2.5mm}
\parbox{37mm}{\begin{fmfgraph*}(100,60) \fmfleft{l} \fmfright{r} \fmf{plain_arrow}{l,r} \end{fmfgraph*}} $\times \Bigg($ \parbox{20mm}{\begin{fmfgraph*}(100,60) \fmfleft{l} \fmfright{r}
\fmf{plain_arrow,right}{l,v1,l}
%\fmf{plain,right}{l,v1}
\fmf{wiggly}{r,v1} \fmfv{d.sh=cross,d.f=empty,d.si=8thin}{v1}
\fmfv{d.sh=square,d.f=full,d.si=6thin}{r}
\end{fmfgraph*}} \hspace{2cm}
+ \hspace{0.5cm} \parbox{20mm}{\begin{fmfgraph*}(100,60) \fmfleft{l}
\fmfright{r} \fmf{plain_arrow,right}{l,v1,l}
%\fmf{plain,right}{l,v1}
\fmf{wiggly}{r,v1} \fmfv{d.sh=cross,d.f=empty,d.si=8thin}{l}
\fmfv{d.sh=square,d.f=full,d.si=6thin}{r}
\end{fmfgraph*}}\hspace{1.6cm}$\Bigg) \quad+$\\
\begin{fmfgraph*}(100,60)
\fmfleft{l} \fmfright{r} \fmf{plain_arrow}{l,v1}
\fmf{plain_arrow}{v1,r} \fmfv{d.sh=cross,d.f=empty,d.si=8thin}{v1}
\end{fmfgraph*}
\end{fmffile}
\caption{Diagrams contributing to $U(1)$ charge
induced}\label{diagU1}
\end{figure}
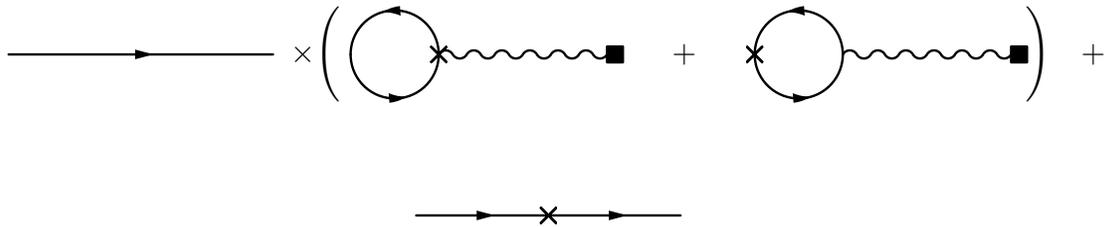 Thus, we can distinguish two contributions: the disconnected
one, $\rho^1(\vec{q})$, coming from the first line in Fig.
\ref{diagU1}, and the connected one, $\rho^2(\vec{q})$, coming from
the second line. We note that the scalar loops contributing to
$\rho^1(\vec{q})$ are precisely the same as those contributing to
the self-energy of $A_{\mu}$ field, thus, \begin{equation}
\rho^1(\vec{q}) = -\int \frac{d q_{\tau}}{2 \pi} K_{0 \nu}(\vec{q},
q_{\tau}) {\cal D}_{\nu \lambda}(\vec{q}, q_{\tau}) J^{{\rm
ext}}_{\lambda}(\vec{q},q_{\tau})\end{equation} Now,
\begin{equation} K_{\mu \nu} (q) {\cal D}_{\nu \lambda}(q) =
\delta_{\mu \lambda} - \frac{q_{\mu} q_{\lambda}}{q^2}\end{equation}
Thus,
\begin{equation} \rho^1(\vec{q}) = - \int \frac{d q_{\tau}}{2 \pi}
\frac{\vec{q}^2}{{q_{\tau}}^2 + \vec{q}^2} J^{{\rm
ext}}(\vec{q},q_{\tau})\end{equation} Noting, $J^{{\rm
ext}}(\vec{q}, q_{\tau}) \stackrel{{\cal T} \to  \infty}{\to} 2 \pi
\delta(q_{\tau})$,
\begin{equation} \rho^1(\vec{q}) = - (1 - \delta_{\vec{q},0})\end{equation} For the
$\vec{q} = 0$ part, the order of the limits $\vec{q} \to 0$,
$q_{\tau} \to 0$ is very important. In our finite system the
$\vec{q} = 0$ mode is isolated, and, moreover, in our present
treatment the Wilson line is of finite length, so we must take the
$\vec{q} = 0$ limit first and then $q_{\tau} \to 0$. Hence,
$\rho^1(\vec{q} = 0) = 0$. This is not surprising. In perturbation
theory, we start with the vacuum which has charge 0. Unless we
manually project the system into a finite charge subspace (as we do
in our treatment by acting on the vacuum with $z$, $z^{\dagger}$
operators), we will never be able to see global screening of charge.
Since the diagrams contributing to $\rho^1(\vec{q})$ are
disconnected from the external $z$ line, $\rho^1(\vec{q} = 0) = 0$.

Now, the connected contribution, simply gives the charge density of
one spinon in the $\vec{k} = 0$ state, \begin{equation}
\rho^2({\vec{q}}) = -\delta_{\vec{q},0}\end{equation} Putting the
two contributions together, \begin{equation} \rho(\vec{q}) =
1\end{equation}
\begin{equation} \rho(\vec{x}) = \delta^2(\vec{x})\end{equation} as expected
by equations of motion (\ref{screen}).

Thus, we have been able to check exact screening of external charge,
which follows from equation of motion (\ref{screen}), to leading
order in $1/N$. We see that local and global parts  of the screening
charge come from very different Feynman diagrams.

\end{document}